\documentclass[journal=jpcafh,manuscript=article]{achemso}
\usepackage{amsmath}
\usepackage{amssymb}
\usepackage[normalem]{ulem}
\usepackage[table]{xcolor}
\usepackage[numbers,super]{natbib}
\usepackage{soul}

\usepackage{xr}
\usepackage{graphicx}
\usepackage{dcolumn}
\usepackage{bm}
\usepackage{xcolor}
\newcommand{\bs}{\ensuremath{\mathbf{s}}}
\newcommand{\br}{\ensuremath{\mathbf{r}}}
\newcommand{\rev}[1]{{\color{black}{#1}}}

\makeatletter
\newcommand*{\addFileDependency}[1]{
  \typeout{(#1)}
  \@addtofilelist{#1}
  \IfFileExists{#1}{}{\typeout{No file #1.}}
}

\makeatother
\newcommand*{\myexternaldocument}[1]{%
    \externaldocument{#1}%
    \addFileDependency{#1.tex}%
    \addFileDependency{#1.aux}%
}
\myexternaldocument{SI}

\title{Iterative unbiasing of quasi-equilibrium sampling}

\author{F. Giberti}
\affiliation{Laboratory of Computational Science and Modeling, Institute of Materials, {\'E}cole Polytechnique F{\'e}d{\'e}rale de Lausanne, 1015 Lausanne, Switzerland}%
\email{federico.giberti@epfl.ch}

\author{B. Cheng}
\affiliation{Trinity College, University of Cambridge, Cambridge CB2 1TQ, United Kingdom}%

\author{G. A. Tribello}
\affiliation{Atomistic Simulation Centre, School of Mathematics and Physics, Queen's University Belfast, Belfast, BT14 7EN, United Kingdom}

\author{M. Ceriotti}%
\affiliation{Laboratory of Computational Science and Modeling, Institute of Materials, {\'E}cole Polytechnique F{\'e}d{\'e}rale de Lausanne, 1015 Lausanne, Switzerland}%
\email{michele.ceriotti@epfl.ch}


\date{\today}
\begin{document}

\begin{abstract}
Atomistic modelling of phase transitions, chemical reactions, or other rare events that involve overcoming high free energy barriers usually entails prohibitively long simulation times.
Introducing a bias potential as a function of an appropriately-chosen set of collective variables can significantly accelerate the exploration of phase space, albeit at the price of distorting the distribution of microstates.
Efficient re-weighting to recover the unbiased distribution can be nontrivial when employing adaptive sampling techniques such as Metadynamics, Variationally Enhanced Sampling or Parallel Bias Metadynamics, in which the system evolves in a quasi-equilibrium manner under a time-dependent bias.
We introduce an iterative unbiasing scheme that makes efficient use of all the trajectory data, and
that does not require the distribution to be evaluated on a grid. The method can thus be used even when the bias has a high dimensionality.
We benchmark this approach against some of the existing schemes, on models systems with different complexities and dimensionalities.
\end{abstract}

\maketitle

\section{\label{sec:introduction}Introduction}
Enhanced Sampling (ES) methods are a cornerstone of molecular modelling,
as they facilitate the investigation of activated processes in chemistry and physics, such as
the folding of macromolecules~\cite{pietrucci2009substrate}, phase transitions in liquids and solids\cite{palmer2014metastable,khaliullin2011nucleation,angi+10prb} as well as chemical reactions in gases and solution\cite{fleming2016new,pietrucci2015formamide}.
ES methods increase the explorative power of Molecular Dynamics (MD) or Monte Carlo by increasing the frequency with which low probability states are visited in simulations.
An early ES method, umbrella sampling\cite{torrie1977nonphysical,torrie1977monte},
applies a static bias potential $V(\bs(\br))$, to confine the system to high free energy states.
Since then, methods using adaptive and time-dependent bias potentials $V(\bs(\br),t)$ have been developed and employed.  In all these methods,
the bias potential is a function of the Collective Variables (CVs), $\bs$, that capture the slow degrees of freedom for the activated process, and which are in turn functions of the atomic coordinates \br .
Notable examples of adaptive ES include Metadynamics, Well-Tempered Metadynamics \cite{laio2002escaping,barducci2008well}, Bias Exchange\cite{piana2007bias},\rev{Parallel Bias Metadynamics\cite{pfaendtner2015efficient}}, Variationally Enhanced Sampling \cite{valsson2014variational}, Basis Function Sampling\cite{whitmer2014basis}, and Artificial Neural Networks\cite{sidky2018learning}.

Generally speaking, ES schemes alter the probability density function for the microstates $P(\br,t)$, and make it deviate from the equilibrium Boltzmann distribution $P(\br)$. Calculating thermodynamic observables, however, requires the equilibrium probability distribution $P(\br)$.
To recover the equilibrium distribution from biased  simulations, it is necessary to reweight each microstate sampled in the MD trajectory in accordance with the bias potential $V(\bs(\br),t)$ acting upon it.
Many different reweighting schemes have been suggested in the past\cite{tiwary2014time,bonomi2009reconstructing,tiana2008estimation},
and they all rely on computing a time-dependent offset $c(t)$ that measures the average accumulated bias over the entire CV space.
However, the analytical solution for $c(t)$ is unknown, and all previous formulations depend on numerically integrating the bias over the CV space.
Performing this integration  has a few drawbacks, as i) the results of the reweighting procedure depends on the grid as well as on the kernel parameters used to calculate the density distribution on the grid; and ii) the computational cost of the methods scales exponentially with the dimensionality of the CV space and the grid density.
Although historically the number of CVs is rarely higher than 3, a number of recent and novel techniques have been suggested to sample more complex phenomena using a larger number of CVs~\cite{awasthi2017exploring,tribello2012using,fu2018lifting}. Reweighting the trajectories generated using these frameworks would be challenging using the methods that are currently available.
To address these issues, we therefore introduce  Iterative Trajectory Reweighting (ITRE), an accurate and grid-independent reweighting algorithm for adaptive ES simulations.
We benchmark our method against the one implemented in PLUMED-2.0\cite{tribello2014plumed} introduced by Vallson \textit{et al.}\cite{valsson2016enhancing} on a few model applications,
including Well-Tempered Metadynamics (WTM)\cite{barducci2008well} calculations for one-particle Langevin systems of different dimensions,
and WTM simulations of Lennard-Jones clusters.
We illustrate that ITRE is on par with state-of-the-art reweighting methods in accuracy, but leads to faster convergence of statistical averages, which we trace to the way different portions of a trajectory contribute to the overall average.

\section{\label{sec:method}Methods}

\subsection{The problem of reweighting}

In an MD simulation performed at constant inverse temperature $\beta$, based of a potential energy surface $U(\br)$, the equilibrium distribution of positions follows the Boltzmann weight $P(\br)\propto e^{-\beta U(\br)}$. Provided that sampling is ergodic, the expectation value of any observable can thus be computed by simply averaging over the configurations encountered along the trajectory.
In order to obtain more representative sampling of high-energy configurations -- such as those that occur during a chemical reaction or a phase transition -- it is often advantageous to introduce an artificial biasing potential $V(\br)$.
This bias is often defined as a function of a small number or collective variables $\bs(\br)$, that are a function of the atomic coordinates and that are the basis for a coarse-grained representation of the system. Although $\bs(\br)$ can be chosen in many ways,
these collective variables should characterize the minima and the transitions states for the reactions that one wants to study~\cite{chen2019nonlinear,mendels2018collective,chen2018collective}.
The presence of the bias distorts the equilibrium probability distribution. However, averages for an observable $O(\br)$ that are consistent with the unbiased Boltzmann distribution can be obtained with a reweighting procedure~\cite{torrie1977nonphysical}, that is
\begin{equation}
\left<O\right>_{\beta,U}  = \frac{\left<O(\br) e^{\beta V(\bs(\br))}\right>_{\beta,U+V} }{\left<e^{\beta V(\bs(\br))}\right>_{\beta,U+V}},
\end{equation}
where $\left<\cdot\right>_{x,y}$ indicates an average computed while sampling the total potential $y$ at inverse temperature $\beta$.

If, instead, the external potential biasing the simulation is time dependent, the resulting trajectory does not sample a well-defined stationary distribution. Provided that the change in external potential is sufficiently slow, however, one can assume that the trajectory evolves in a quasi-equilibrium fashion, and that at each time the biased distribution $\hat{P}$ is consistent with the sum of the physical and bias potentials, i.e.
\begin{equation}
\hat{P}(\br,t) \propto e^{ -\beta\left[V(\bs(\br),t)+U(\br)\right]} = P(\br) e^{-\beta V(\bs(\br),t)}.\label{eq:pp-prop}
\end{equation}
In order to compute an unbiased average from a simulation with a time dependent bias potential, one must, therefore, combine different portions of the trajectory, each of which is consistent with the instantaneous value of $V(\bs,t)$ acted upon them. To do so, the probabilities computed at different times must be compared on the same footings, which means that it is not sufficient to define the probability distribution modulo an unknowns normalization constant as in Eq.~\eqref{eq:pp-prop}.
Usually, the missing normalization is expressed as a time-dependent offset $c(t)$, i.e.
\begin{equation}
    \hat{P}(\br,t) = P(\br)\ e^{-\beta [V(\bs(\br),t) - c(t)]},
\label{eq:generalP}
\end{equation}
in which
\begin{equation}
 e^{-\beta c(t)} = \frac{\int\ d \bs\ P(\bs)\ e^{-\beta V(\bs,t)}}{\int d \bs\ P(\bs)},
\label{eq:probab_omega}
\end{equation}
corresponds to the ratio of the partition functions of the biased and unbiased distributions.

Over the course of the years, several procedures have been proposed to calculate the time-dependent offset $c(t)$, and these have recently been compared by Gimondi \textit{et al.} \cite{gimondi2018building}. Even though such procedures have been given different formal justifications, they can all be understood within a single framework. For many adaptive ES methods, the quasi-equilibrium regime that underlies Eqns.~\eqref{eq:generalP} and \eqref{eq:probab_omega} is achieved only after the minima in the CV space have all been sampled, and the bias is quasi-stationary.
Given that, for a real system, one does not know a-priori when quasi-stationary conditions are achieved, and in order to utilize the early portions of the trajectory, one can instead define a time-dependent \emph{unbiased} $P(\bs,t)$, and exploit the fact that in early stages of an ES trajectory equilibration is achieved locally even without an ergodic sampling of the whole space\rev{, as recently also suggested by Marinova \textit{et al.} \cite{marinova2019time}}.
This leads to a general expression for $c(t)$
\begin{equation}
 e^{-\beta c(t)} = \frac{\int\ d \bs\ P(\bs,t)\ e^{-\beta V(\bs,t)}}{\int d \bs\ P(\bs,t)},
\label{eq:out_equilibrium_c_t}
\end{equation}
with different approaches for estimating $c(t)$ emerging as a consequence of taking different ansatzes for $P(\bs,t)$.
For example, Bonomi \textit{et al.} \cite{bonomi2009reconstructing} suggested propagating $P(s,t)$  rather than estimating $c(t)$ directly
\begin{equation}
P(\bs,t+\delta t) = e^{-\beta[\dot{V}(\bs,t)-\dot{c}(t)] \delta t}P(\bs,t),
\end{equation}
where $\dot{c}(t)$ can be calculated as
\begin{equation}
\dot{c}(t) = \int d \bs \ \dot{V}(\bs,t)P(\bs,t).
\end{equation}

A more recent method by Tiwary and Parrinello \cite{tiwary2014time} provided an analytic formulation for $P(\bs,t)$ in the case of Metadynamics\cite{laio2002escaping}. Rather than
reproducing their derivation,
we note that the same result can be obtained using the ansatz
\begin{equation}
P(\bs,t)=e^{\beta V(\bs,t)} g(\bs-\bs(t)),
\end{equation}
where $g(\bs-\bs(t))$ is the repulsive Gaussian deposited at time $t$ with variance $\sigma$.
Consequently, $c(t)$ can be expressed as
\begin{equation}
e^{\beta c(t)}=\frac{1}{(2\pi)^{d/2}\textrm{det}(\sigma)}\int d \bs e^{\beta V(\bs,t)} g(\bs-\bs(t)).
\label{eq:pratyush_c_t}
\end{equation}

For a WTM calculation,  $c(t)$ can also be computed by taking finite differences of the exponent of the bias in two subsequent steps~\cite{tiwary2014time}.
More recently, Valsson \textit{et al.}\cite{valsson2016enhancing} used a different form of $P(\bs,t)$ that is valid for WTM\cite{barducci2008well} only:
\begin{equation}
P(\bs,t)  = e^{\Delta \beta V(\bs,t)},
\label{eqn:valsson_p}
\end{equation}
with $\Delta \beta= [ k_B T(\gamma - 1) ]^{-1}$ where $\gamma$ is the bias-factor in WTM methods, which leads to
\begin{equation}
c_s(t)=\frac{1}{\beta}\log\frac{\int d \bs \ e^{\gamma\Delta \beta V(\bs,t)}}{\int d \bs \ e^{\Delta \beta V(\bs,t)}}.
\label{eq:valsson_c_t}
\end{equation}
Eqn.~\eqref{eqn:valsson_p} and ~\eqref{eq:valsson_c_t} represent the reweighting scheme that is currently implemented in the PLUMED-2.0~\cite{PLUMED} suite, which we will reference as $c_s(t)$, and use as a comparison.

\subsection{Iterative determination of P(\bs,t)}

The choice of $P(\bs,t)$ affects the accuracy of the weights, and also the applicability to different biasing schemes.
By taking Eqn.~\eqref{eq:out_equilibrium_c_t} as an ansatz for $c(t)$, $P(\bs,t)$ can be obtained directly from the biased trajectory, by computing a re-weighted histogram:
\begin{equation}
 P(\bs,t) = \frac{\int_0^t k(\bs-\bs(t'))e^{\beta[V(\bs(t'),t')-c(t')]} dt'}{\int_0^t e^{\beta[V(\bs(t'),t')-c(t')]} dt'},
\label{eq:partial_P_s_t}
\end{equation}
where $k(\bs-\bs(t'))$ is a binning function that is used to obtain a smooth histogram.
The fact that the probability density $P(\bs,t)$ depends on $c(t)$ means that, unlike in the schemes proposed by Tiwary, Valsson or Bonomi, $c(t)$ cannot be calculated on-the-fly.
What we propose is instead to estimate $c(t)$ self-consistently, following an iterative protocol:
\begin{enumerate}
\item Initialize $c(t)$ to zero, and evaluate the probability density function $P(\bs,t)$ with equation \eqref{eq:partial_P_s_t}.
\item With this approximated $P(\bs,t)$, evaluate $c(t)$ using equation \eqref{eq:out_equilibrium_c_t}.
\item Re-evaluate $P(\bs,t)$ with the new $c(t)$ through equation \eqref{eq:partial_P_s_t}.
\item Compare $P(\bs,t)$ between two consecutive iterations and terminate if it converges.
\end{enumerate}
\rev{This iterative approach is reminiscent of that used in the weighted histogram analysis method~\cite{roux1995calculation,kumar1992weighted}. We note however that it is not possible to map the two methods onto one another by slicing the biased trajectory into windows, unless the bias changes slowly enough to fully converge the histogram for each window, which is hardly ever possible for typical history-dependent biasing schemes.
}

ITRE is more general than the previous reweighing schemes, as we made no assumptions on the form of  $V(\bs,t)$, or on the existence of a direct relationship between the bias and the quasi-stationary probability distribution. However, its accuracy still depends on the grid used to calculate $P(\bs,t)$ and on the smoothing of the binning function $k(\bs,\bs(t'))$.
In addition, as was the case for previous approaches, the need to integrate over a discrete grid limits its applicability to low-dimensional problems.

\subsection{Trajectory-based determination of $c(t)$}

We can avoid calculating $P(\bs,t)$ on a grid by substituting equation \eqref{eq:partial_P_s_t} directly into equation \eqref{eq:out_equilibrium_c_t}, and by integrating over $\bs$ considering the limit in which the binning function is an actual Dirac $\delta$ distribution.
In that limit, the self-consistent equation for $c(t)$ can be written as an integral over the trajectory
\begin{equation}
    e^{-\beta c(t)} = \frac{\int_0^t dt'\ e^{\beta[V(\bs(t'),t')-c(t')-V(\bs(t'),t)]}}{\int_0^t dt'\ e^{\beta [V(\bs(t'),t')-c(t')]}}.
\label{eq:time_c_t}
\end{equation}
In this framework $c(t)$ can now be evaluated without
computing the bias function at all grid points in space. Instead the instantaneous value for the bias $V(\bs(t'),t)$ for all the  configurations in the trajectory with $t' \le t$ must be evaluated.
As with the grid-based iterative approach, $c(t)$ can be computed straightforwardly for any form of the bias $V$(\bs(\br),t).
The clear downside of this trajectory-based definition is that the cost of this analysis depends quadratically on the length of the trajectory. Essentially, the trajectory data provides a sparse grid of points on which to estimate the ratio of biased and un-biased partition functions, which becomes advantageous as the dimensionality of the collective variable space increases. As we will show below, the sampling of the trajectory can be made quite sparse without degradation of the reweighting performance, so in practice the cost of the procedure is not an issue.
\rev{The method illustrated in equation \eqref{eq:time_c_t} has been implemented in the the development version of plumed2, available on \url{https://github.com/plumed/plumed2}, SHA-1 b39e0c4df0f229dffdd09b410f7fee169470b15c. The inputs to perform the simulations are deposited on Plumed Nest, the public repository of the PLUMED consortium \cite{bonomi2019promoting}, as plumID:19.078.}

\section{\label{sec:benchmarks}Benchmark systems}

To benchmark the performance of ITRE against the method that was implemented in PLUMED by Valsson \textit{et al.}, we performed WTM simulations of a Langevin particle in an analytic potential with dimensionality $D=2$, 3 and 6 using Plumed-2.0 \cite{PLUMED}.
In addition, we simulated the solid-solid phase transition of a Lennard-Jones 38 cluster (LJ-38) with the LAMMPS code and Plumed-2.0 \cite{PLUMED,plim95jcp} using WTM calculations. In both these problems it would be easy to generate starting conditions that are consistent with the equilibrium Boltzmann distribution.
We chose, however, to start all trajectories from the global minimum on the potential energy surface, to mimic realistic conditions in which the starting configurations (e.g. determined by experimental data) introduce a bias. We will assess how different reweighting schemes deal with this bias, as well as with the statistical uncertainty in the estimate of the FES.

\section{\label{sec:results}Results}

\subsection{\label{subsec:convergence}Convergence of $c_I(t)$}

We begin our discussion by investigating the convergence of Eqn.~\eqref{eq:time_c_t} in terms of the number of iterations $I$, and how robust it is to an increase of the spacing $\tau$ between the time steps at which the integral is evaluated.
To perform these tests, we used a WTM simulation exploring the 2D potential with three minima illustrated in Fig.~S1, using both $x$ and $y$ as CVs in the WTM calculation.
Since the cost of evaluating $c(t)$ scales quadratically with the simulation length, we checked if $c_I(t)$ can be estimated sparsely during the trajectory, which can be useful when analyzing very long trajectories. Since $c(t)$ is a slow-changing function,
estimating it every $\tau$ steps should not  affect the accuracy of the probability estimation.
As a rule of thumb, $\tau$ should be smaller than the time scale over which the system hops from one free energy minimum to another.

\begin{figure}
    \centering
    \includegraphics[width=\textwidth]{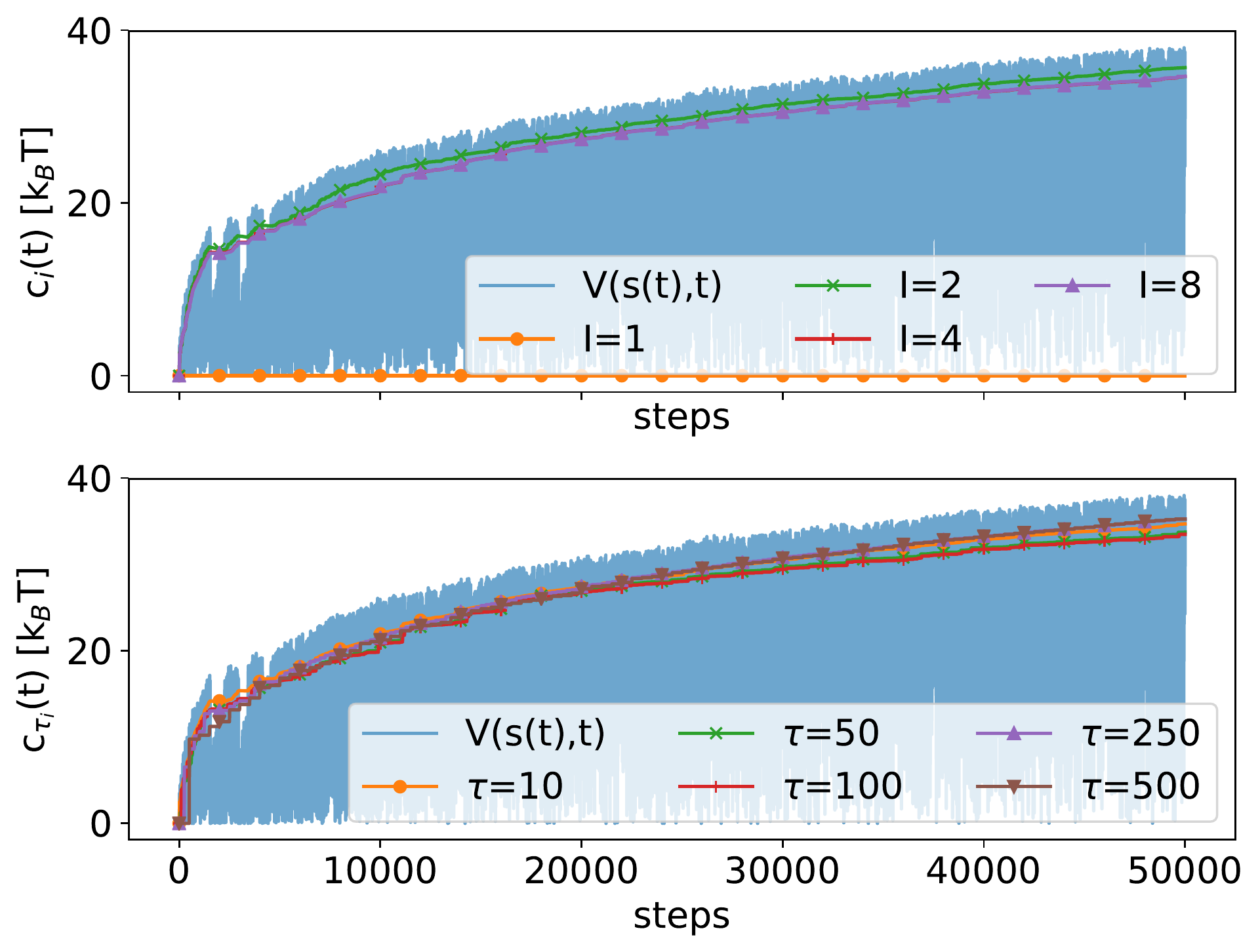}
    \caption{Convergence of $c_I(t)$ as a function of the number or iteration $I$ (upper panel) or stride $\tau$ (lower panel) for a one-dimensional FES of a 2D system. The small changes in $c_I(t)$ that occur between the 2nd and 3rd iteration will have a tiny effect on the recovered FES so the calculation is effectively converged by this point. ITRE seems also to be slightly affected by the choice of $\tau$ as can be seen in the lower panel. To decide which stride is the best to use, it is better to compare different strides and choose the one that minimize the error in the FES. In this case, we decided to use a stride of 100 Gaussian kernels deposition. A figure representing the absolute error on the FES is reported in the Supplementary Information in Fig.~S3.
    }
    \label{fig:n_iter}
\end{figure}

The results presented in the upper panel of Fig.~\ref{fig:n_iter} show that $c_I(t)$ converges after a small number of iterations - barely changing after the first two iterations. Thus, any unbiased property can be safely calculated from the 3rd iterations on. $c_I(t)$ also appears to be robust with respect to the stride parameter $\tau$. As illustrated in the lower panel of Fig.~\ref{fig:n_iter}, changing the $\tau$ value has a minimal effect on $c_I(t)$. In this case, we found that the maximum stride that should be used is approximately 100 times the Gaussian deposition stride. We recommend testing the convergence of the FES for both $I$ and $\tau$, to guarantee a high level of accuracy in the FES evaluation. We illustrate how the error on the FES changes as a function of $I$ and $\tau$ in the SI for the interested reader.

\begin{figure}
    \centering
    \includegraphics[width=0.65\textwidth]{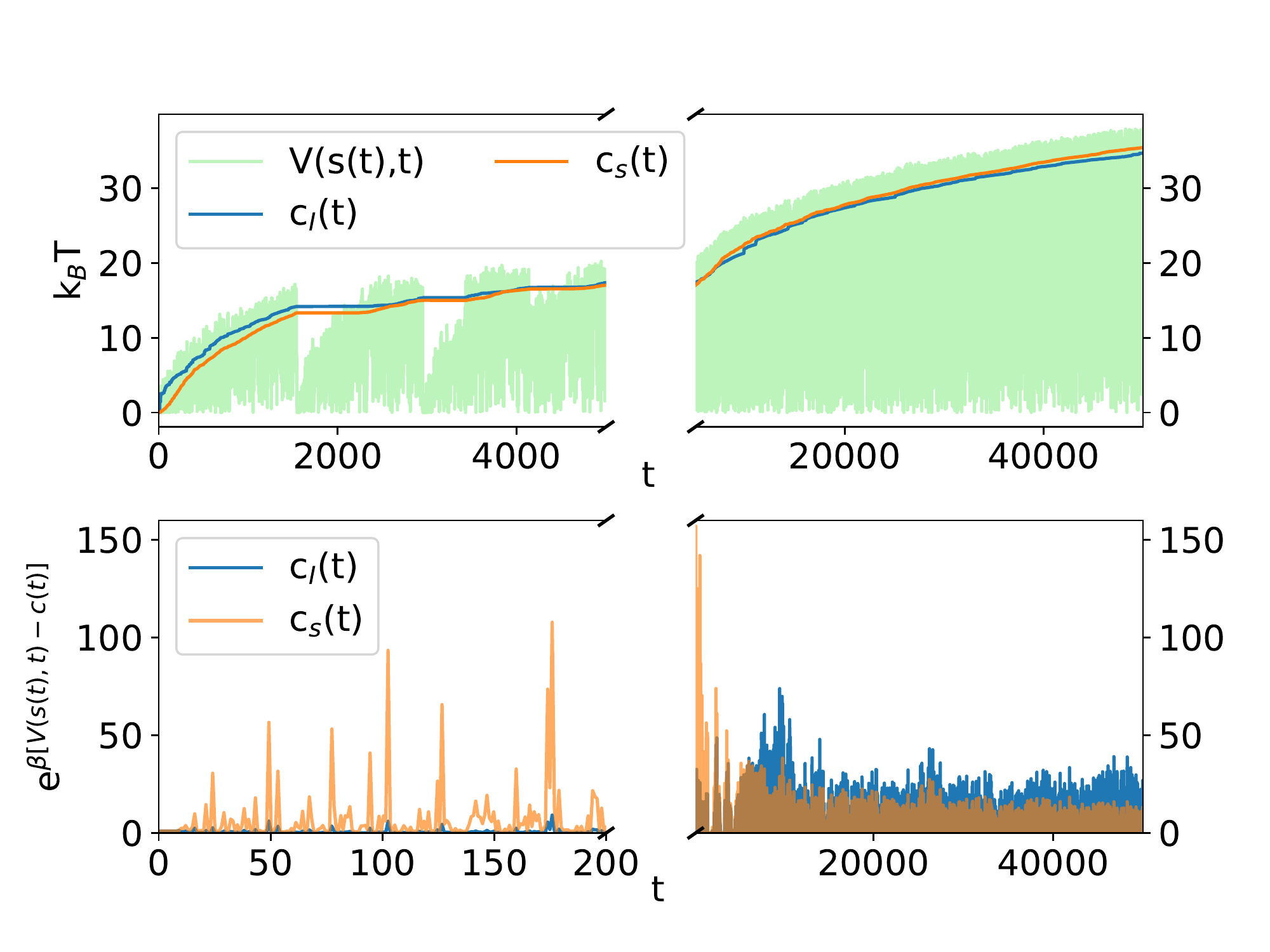}
    \caption{Comparison between $c_I(t)$ and $c_s(t)$ and the corresponding weights for Gaussians with small widths. Despite the small differences, which are barely noticeable in the upper panel, since the weights depend on the exponent of $\beta [V(\bs(t),t) - c(t)]$, there is a large difference in the weights of each microstate, as illustrated in the lower panel. As explained in the text, $c_s(t)$ assign substantial weights to early microstates, while $c_I(t)$ assigns more weight to the end of the trajectory and thus provides overall weights that are more homogeneous.}
    \label{fig:weights}
\end{figure}

Having tested the robustness of our scheme, we compared the behavior of $c_I(t)$ and $c_s(t)$.
The upper panel of Fig.~\ref{fig:weights} shows that $c_I(t)$ increases monotonically inside the basin of a minimum and decreases slightly when a transition occurs, as previously noticed by Tiwary \textit{et al.}. However, there are quantitative differences between ITRE and $c_s(t)$. In particular, during the early stages of the trajectory $c_I(t)$ is higher than $c_s(t)$, while in the later parts of the trajectory the order of these two quantities reverses.
Since the weights depend on the difference $V(\bs(t'),t') -  c(t)$, $c_s(t)$ weights the microstates at the beginning of the calculations more, which is unfortunate as these configurations are out of the equilibrium since the bias is rapidly increasing. $c_I(t)$ behaves in direct contrast to $c_s(t)$ and assigns the lowest weights to the early microstates and thus weights microstates at the end of the trajectory, for which the bias is almost stationary for WTM calculations, more.
Usually, the over counting of the early stages in WTM calculations is avoided by discarding the initial part of the trajectory, which is affected by the choice of starting conditions, and might therefore introduce an unknown bias in the estimate of the FES.
The fact that $c_I(t)$ naturally yields smaller weights in the early stages of the calculation, suggests that ITRE should be less affected by the initial conditions and early microstates, even in cases in which it is not obvious how to choose what portion of the trajectory should be discarded.
Although weakly dependent on the subsampling of the trajectory, we found the weights calculated with ITRE are susceptible to the width of the Gaussians deposited during the WTM calculations, with wider Gaussians leading to large fluctuations in the weight and a degradation of the statistical efficiency of reweighting~\cite{ceri+12prsa}. This behavior arises because Gaussian kernels have infinite support, so even configurations that are far apart in CV space and which belong to different minima will influence each other if the kernel is not truncated.
A comparison of the weight $e^{\beta [V(\bs(t'),t') -  c(t)]}$ is reported in the lower panel of Fig.~\ref{fig:weights} for Gaussians with $\sigma \approx0.25$ times the width of the minima and Fig.~S4 for Gaussians with $\sigma$ values that are the same size of the minima for the system.

\subsection{Convergence of the FES with different reweighing schemes}

As another test, we performed a statistical analysis on $c_s(t)$ and $c_I(t)$ for a set of 32 calculations in a system with $D=3$ and where all three variables were used as CVs in the WTM calculations. Slightly different initial coordinates near the minimum at the origin were used for the initial position of the particle in the various simulations. \rev{For all these cases, a stride of $\tau=$10 Gaussian deposition steps was used to evaluate both $c_s(t)$ as well as $c_I(t)$, as we found it provides sufficient accuracy in the calculation of the FES.}

We computed the one and two-dimensional projections of the probability density $P(\bs)$ and the associated free energy $F(\bs)$.
For this system with an analytic potential function,
the probability density function $\tilde{P}(\bs)$ can be calculated exactly as
\begin{equation}
    \tilde{P}(\bs) = \frac{\int e^{-\beta U(\mathbf{r})} \delta(\bs-\bs(\mathbf{r})) d\mathbf{r}}{\int e^{-\beta U(\mathbf{r})} d\mathbf{r}},
\end{equation}
so the corresponding FES is
\begin{equation}
    \tilde{F}(\bs) = - \beta^{-1} \log \tilde{P}(\bs).
\end{equation}
We then calculated the Kullback-Leibler (D$_{KL}$) divergence\cite{kullback1951information} between the analytic  $\tilde{P}(\bs)$ and the reweighted one $P(\bs)$ using ITRE and $c_s(t)$, using
\begin{equation}
    D_{KL} = \int \tilde{P}(\bs) \log \left ( \frac{\tilde{P}(\bs)}{P(\bs)} \right ) d \bs.
\end{equation}
We would like to highlight that the choice of $D_{KL}$ as a metric to compare the accuracy of the FES reconstruction is physically justified, as
\begin{equation}
    D_{KL} = \int \tilde{P}(\bs)\ \beta \left ( F(\bs) - \tilde{F}(\bs) \right ) d \bs,
\end{equation}
meaning that $D_{KL}$ evaluates the difference between the analytic and reconstructed FES weighted by $\tilde{P}(\bs)$. The larger the $D_{KL}$, the more significant is the difference between the two FES.

\begin{figure}
    \includegraphics[width=\textwidth]{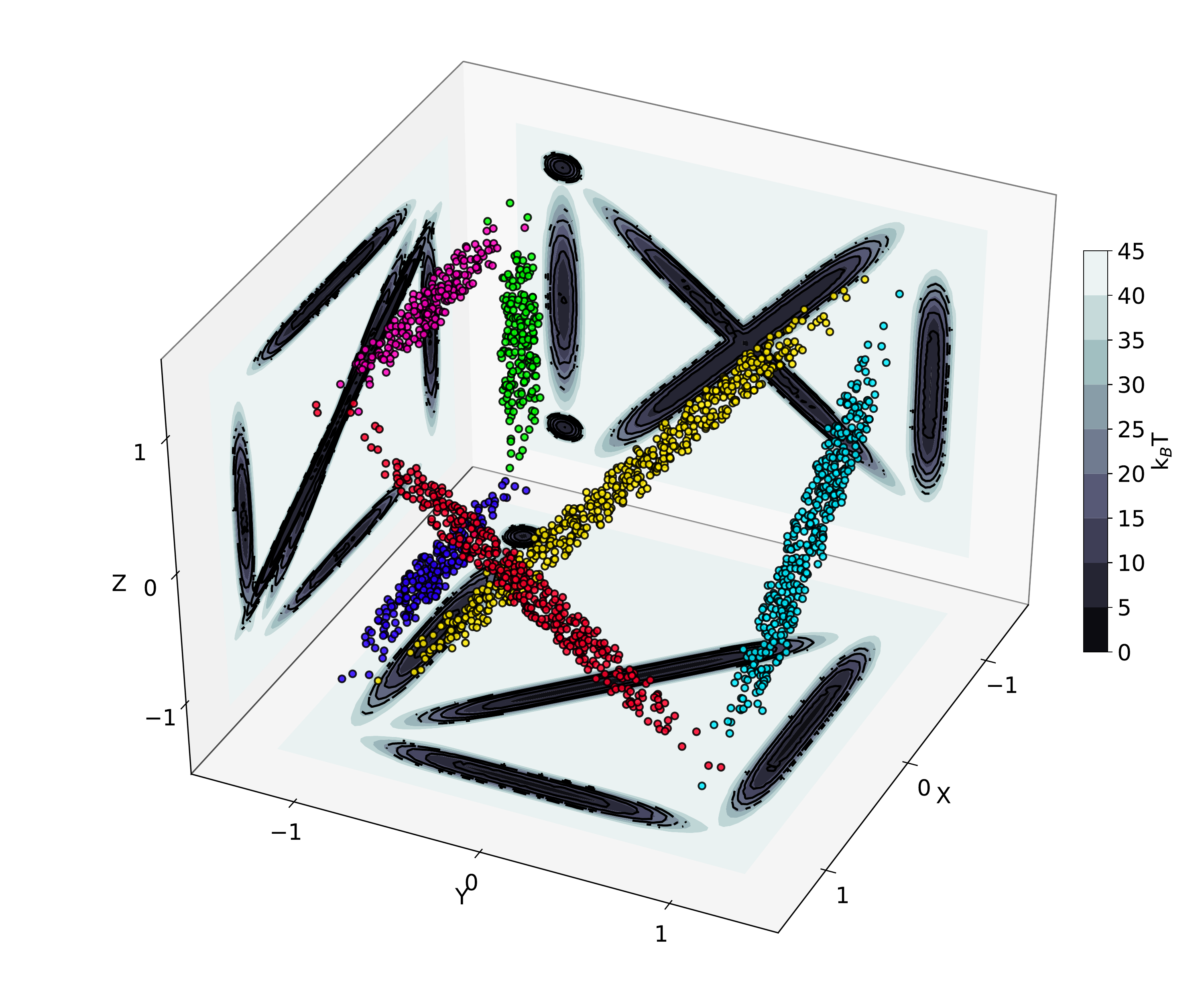}
    \caption{Illustration of the 3D system used to test the behavior of ITRE. The six different minima in this energy landscape are highlighted in different colors.  Furthermore,  the FES in the three 2D planes, $(x,y)$, $(x,z)$, and $(y,z)$ are also reported. The analytic FES is reported in a gray-scale, while the reconstructed FES is drawn using continuous black lines.}
    \label{fig:3d_system}
\end{figure}

\begin{figure}
    \includegraphics[width=0.8\textwidth]{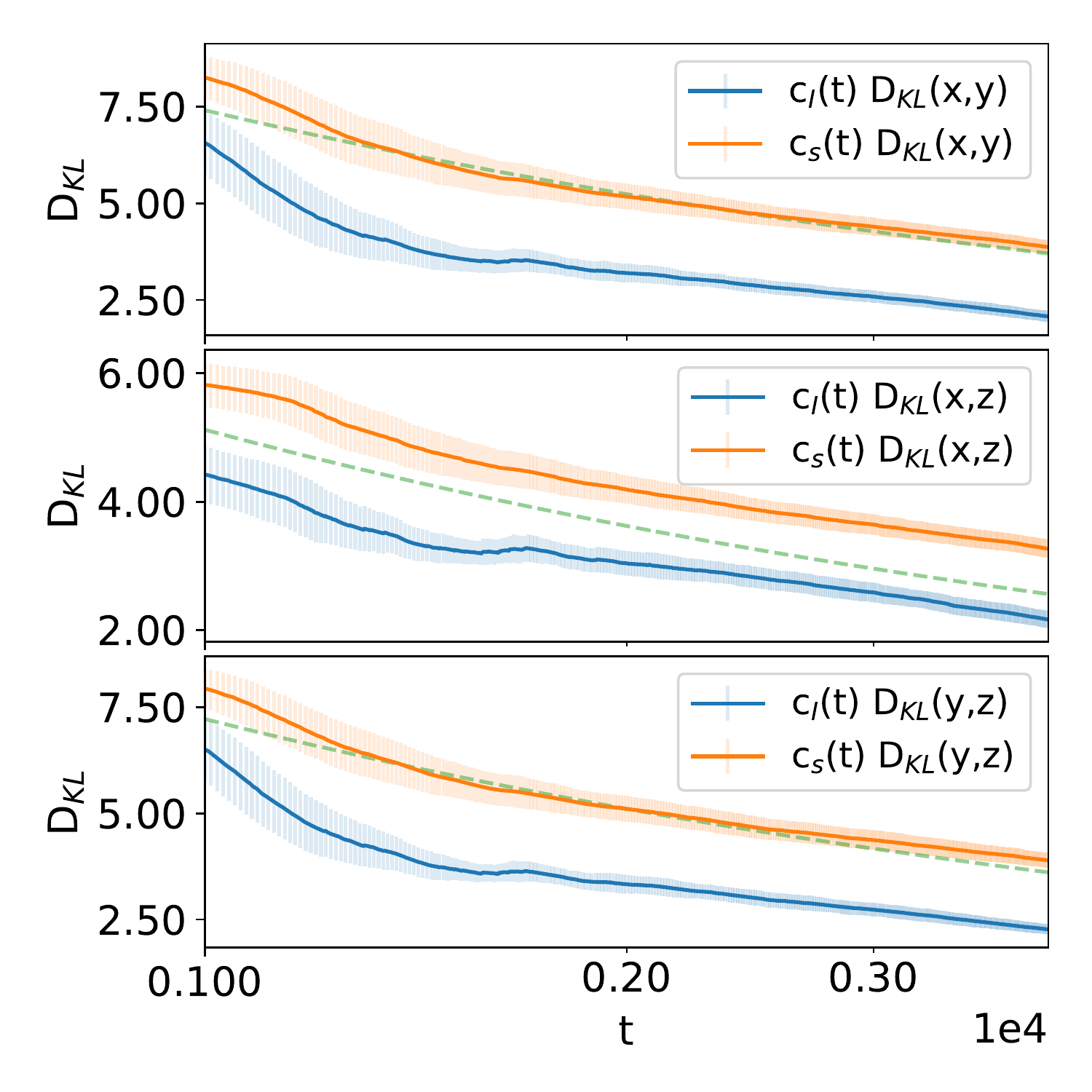}
    \caption{From top to bottom, the accuracy in the calculation of $P(\bs)$ using different reweighting methods measured using the KL divergence, $D_{KL}$, for the $(x,y)$,
    $(x,z)$ and $(y,z)$ planes respectively. As can be seen ITRE converges faster than $c_s(t)$\rev{in all cases. The green dashed line represent the $1/\sqrt{t}$ behavior that the KL divergence should have in the long time limit.}}
    \label{fig:KL_divergence_3d}
\end{figure}

For brevity, we report the mono-dimensional FES and $D_{KL}$ in the SI, while here we discuss only the bi-dimensional FES and $D_{KL}$ of the $D=3$ system. The three FES, as well as a sketch of the 3D potential with the six different minima colored differently,  is reported in Fig.~\ref{fig:3d_system}. From the plot of $D_{KL}$ in Fig.~\ref{fig:KL_divergence_3d} it is clear that ITRE converges faster to the correct probability distribution function for the three cases.
Both $c_s(t)$ and $c_I(t)$ provide the correct estimate of the FES in the long-time limit, but in these relatively short calculations subject to the systematic bias of starting from a non-equilibrium distribution, there is a noticeable difference. The slower convergence of $c_s(t)$ can, in fact, be attributed to the large weights that it assigns to the early stages of the simulations, which requires extended sampling to correct.

\subsection{Applications to high-dimensional sampling}

To further illustrate the flexibility of ITRE for high dimensional systems, we performed WTM simulations for the a Langevin particle in a space with $D=6$. The analytical potential we use contains three interconnected minima, and we perform ES with metadynamics, using both the conventional WTM scheme and the diffusive adaptive Gaussians method~\cite{branduardi2012metadynamics}. All six coordinates were used as CVs in the WTM calculations. In the diffusive adaptive Gaussians scheme, the normal WTM repulsive Gaussian is substituted with a multivariate Gaussian where the elements of the covariance matrix are not defined a-priori but are calculated on the fly during the trajectory. With the diffusive scheme, the values of the CVs and the elements of the covariance matrices are updated using

\begin{equation}
\begin{split}
 \dot{\bar{s_i}}(t) =& \frac{s_i(t)-\bar{s}_i(t)}{\chi},\\
    \dot{\bar{\sigma_{ij}}}^2(t) =& \frac{[s_j(t)-\bar{s}_j(t)][s_i(t)-\bar{s}_i(t)]-\sigma_{ij}^2(t)}{\chi}.
\end{split}
\end{equation}

This formulation is equivalent to an exponential moving average of $s_i$ and $\sigma_{ij}$, where $\chi$ determines the time scale over which the average is performed. The advantage of this method is that the repulsive Gaussians adapt to the local shape of the FES, and thereby introduce a bias that is sufficiently precise to reproduce the FES faithfully, and smooth enough to quickly fill up the free energy minimum, accelerating the exploration of the FES..
This approach is however challenging for reweighting schemes. On one hand, the relationship between bias and free energy is not trivial~\cite{branduardi2012metadynamics}. On the other, the accuracy with which a kernel can be interpolated on the grid depends on the magnitude of the covariance matrix, leading to potential large errors if the Gaussians become too sharp relative to the grid spacing, and eventually to  noise in the bias that affects the statistical efficiency in reconstructing the Boltzmann weights. Besides these issues related to the adaptive Gaussian width, one should note that even when using conventional metadynamics, evaluation of a histogram on a six-dimensional grid is prohibitively expensive in terms of both computational effort and memory footprint.

\begin{figure}
\includegraphics[width=0.8\textwidth]{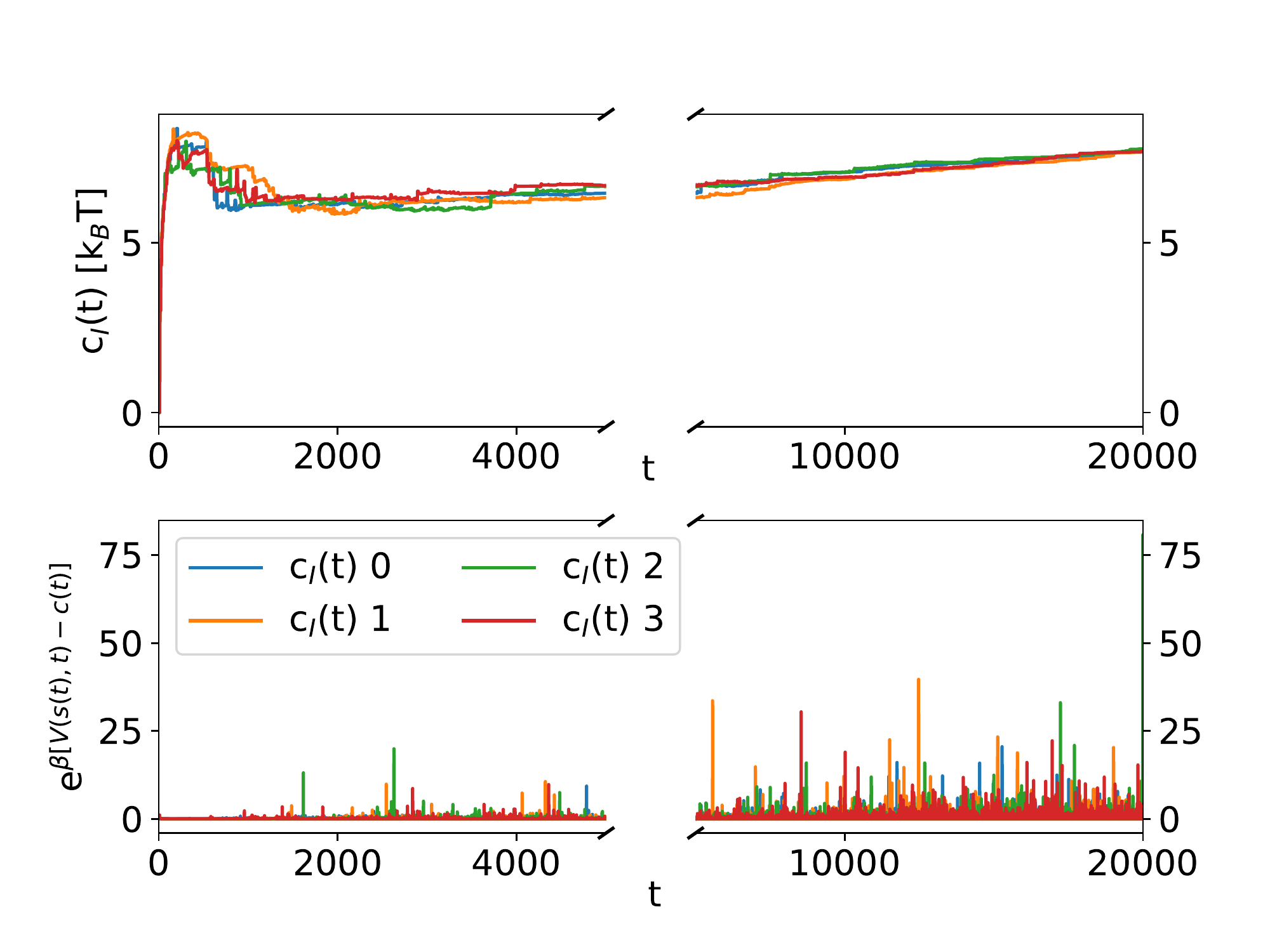}
    \caption{Time behavior of $c_I(t)$ (upper panel) and $e^{\beta [V(s(t),t) - c_I(t)]}$ (lower panel). Some of the weights present very large values due to the overlap of kernels belonging to different minima.}
    \label{fig:6_d_adaptive_weights}
\end{figure}

\begin{figure}
\includegraphics[width=\textwidth]{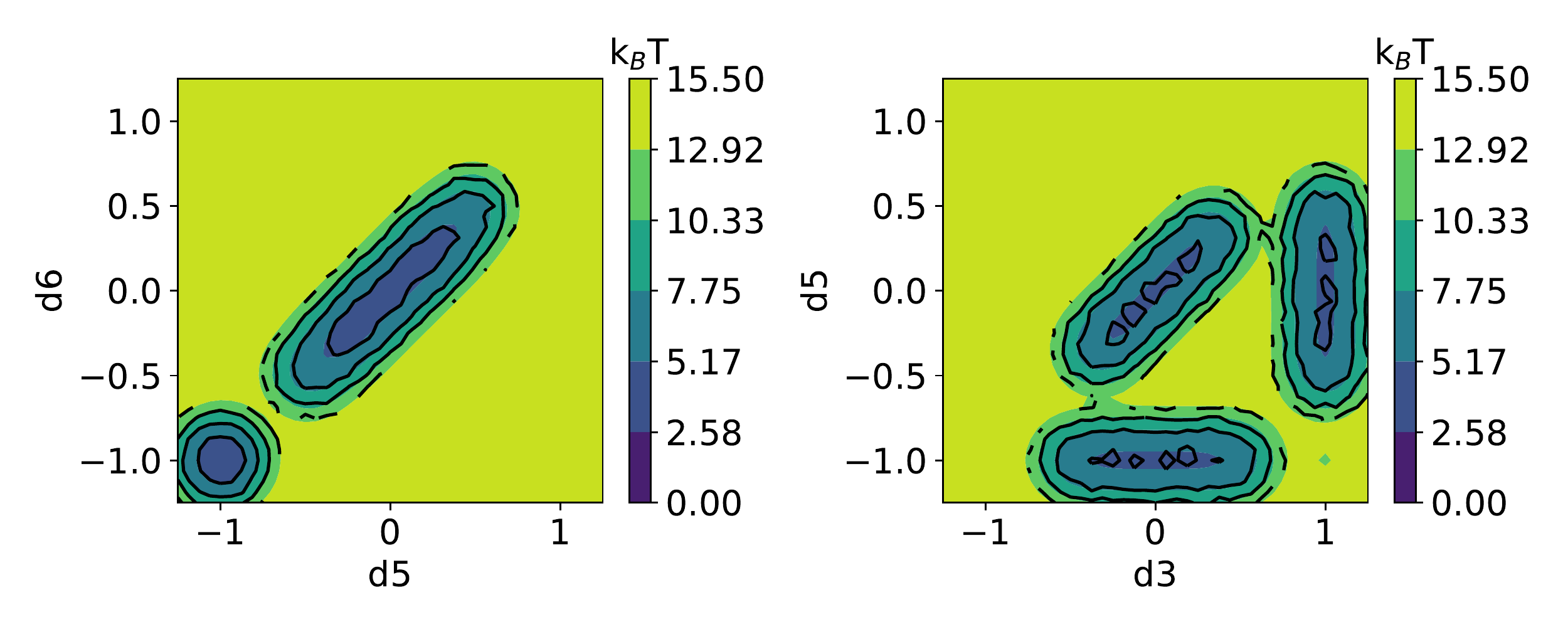}
    \caption{FES for the $D=6$ system calculated as  a function of $d5$ and $d6$, as well as $d3$ and $d5$. The analytic FES is reported in filled contours, while the reweighted FES is the solid black line. A difference plot is reported in the SI.}
    \label{fig:6_d_FES}
\end{figure}

Since ITRE does not depend on a grid, we can easily reweight the bias introduced with adaptive Gaussians and calculate the weights, which we reported in Fig.~\ref{fig:6_d_adaptive_weights}. As shown for lower-dimensional cases, ITRE assigns low weights to the first portion of the trajectory, and higher weights to the late microstates that follow achievement of quasi-equilibrium conditions. However, as can be noticed in the lower panel of Fig.~\ref{fig:6_d_adaptive_weights} some microstates have a substantial higher weights. As explained at the beginning of Sec.~\ref{subsec:convergence}, this is due to the large width assigned to the Gaussian kernels during the simulations, which introduces spurious correlations between microstates belonging to different minima.  The FES obtained with such weights is illustrated in Fig.~\ref{fig:6_d_FES}, showing that even without perfectly smooth weights, ITRE can still produce a  FES that is comparable to the analytic one, with only four trajectories that consist of $\approx$20000 deposited Gaussian. The projection of the FES along individual directions, as well as the FES obtained with constant hill size and the corresponding weights,  are reported in Sec.~S6.

\subsection{Application to Lennard-Jones 38}

The last example that we report here is the FES of an LJ-38 cluster in vacuum at a temperature of T$^*$=0.168, using the two collective variables $n_6$ and $n_8$, which evaluate the number of six-coordinated and eight-coordinated atoms in the cluster. In more details, $n_c$ is expressed as
\begin{equation}
n_{c}=\sum_{i=1}^{N}e^{-\frac{\left(c_{i}-c\right)^{2}}{2\eta^{2}}},
\end{equation}
where the coordination $c_{i}$ for each atom is calculated as a function of the distance $d$ between them,
\begin{equation}
c_{i}=\sum_{j}\mathcal{S}\left(\left|\br_{i}-\br_{j}\right|\right),\qquad\mathcal{S}\left(d\right)=\begin{cases}
0 & d>r_{0}\\
1 & d<r_{1}\\
\left(y-1\right)^{2}\left(2y+1\right) & r_{1}<d<r_{0},\quad y=\frac{d-r_{1}}{r_{0}-r_{1}}
\end{cases}.
\end{equation}
For these simulations, we used the parameters $\eta=0.5$, $r_0=1.5$ and $r_1=1.25$ reduced units, respectively.

\begin{figure}
\includegraphics[width=0.8\textwidth]{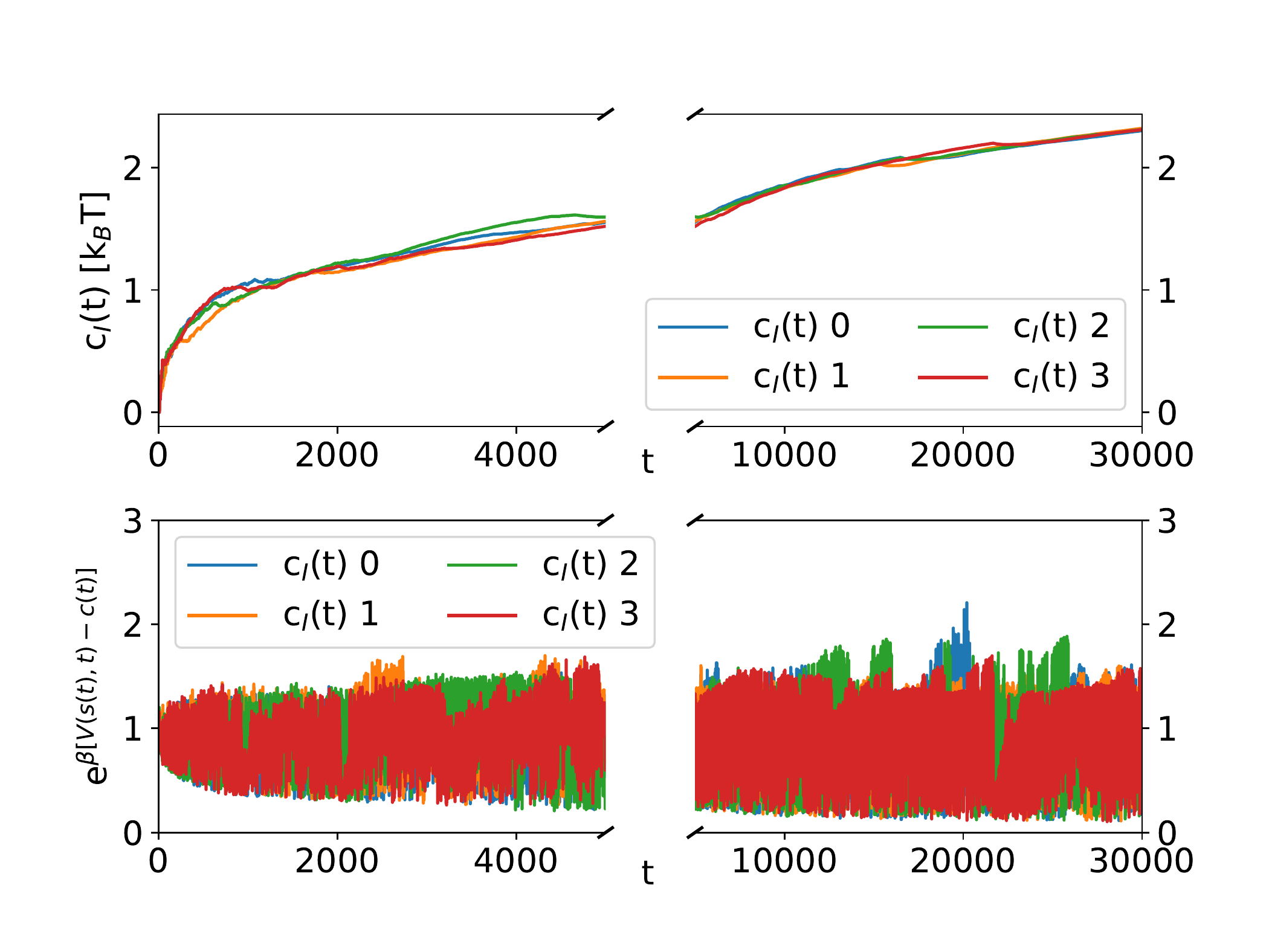}
    \caption{Time-dependent behavior of $c_I(t)$ (upper panel) and $e^{\beta [V(s(t),t) - c_I(t)]}$ (lower panel).For this example, both $c_I(t)$ and the associated weights are very well-behaved.  The resulting FES, reported in Fig.~\ref{fig:LJ-38} is converges smoothly and quickly.}
    \label{fig:LJ-weights}
\end{figure}

\begin{figure}
\includegraphics[width=\textwidth]{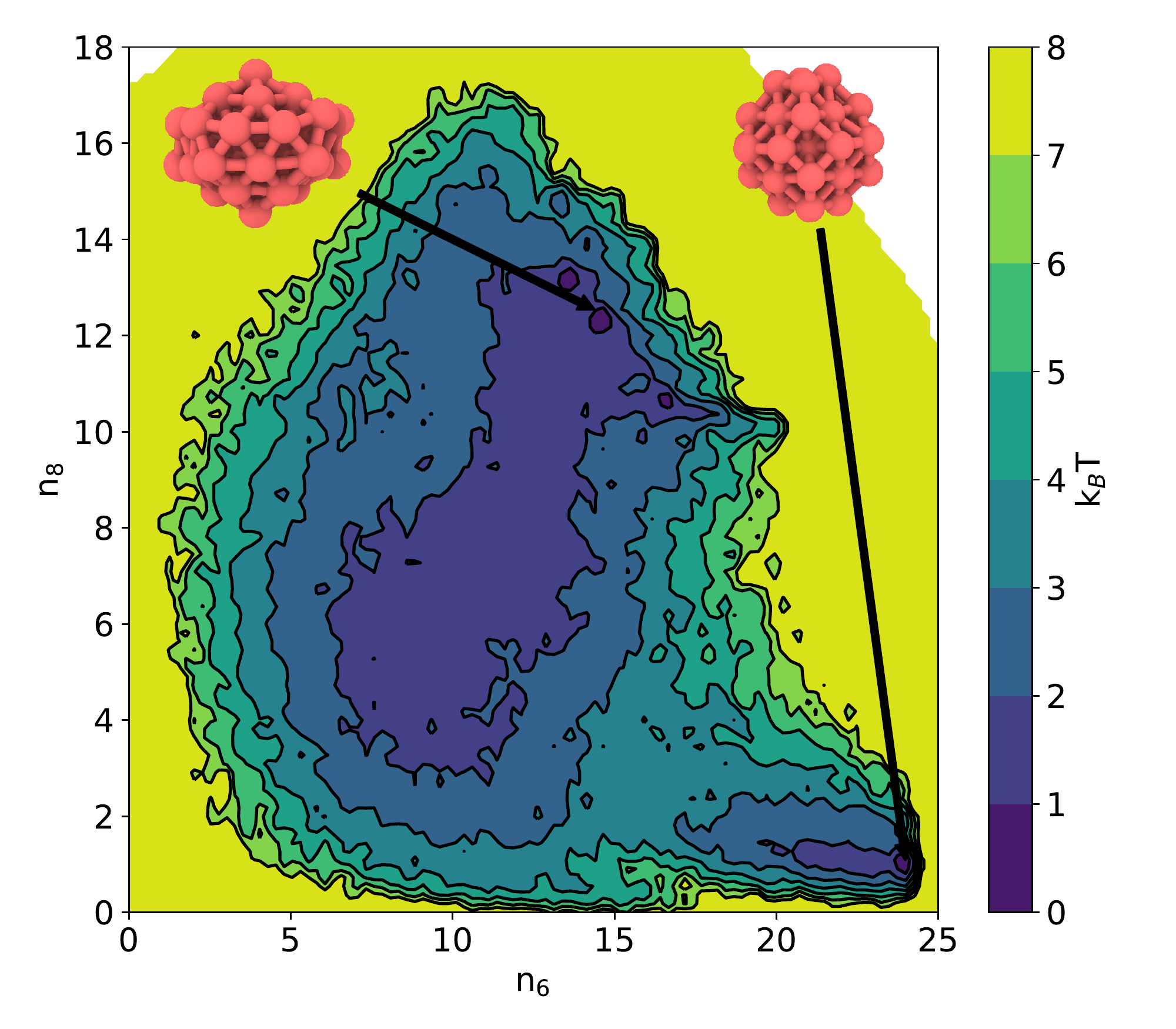}
    \caption{FES for the LJ-38 cluster in vacuum as a function of the CVs $n_6$ and $n_8$. The two most stable structures have also been reported in figures with a ball-and-stick representation, with black arrows pointing at their resepective position on the $n_6$, $n_8$ landscape.}
    \label{fig:LJ-38}
\end{figure}

The LJ-38 is a paradigmatic example of a system presenting an energy landscape with two-minima in strong competition.  For this system the global minimum is entropically disfavoured with respect to the other minimum and these two minima \rev{are separated by a barrier of a few k$_B$T}. The FES is thus simple and yet has the characteristics of more complex systems. The $c_I(t)$ and the weights calculated through ITRE for this example are very well behaved (see  Fig.~\ref{fig:LJ-weights}), showing that application to a realistic sampling problem does not introduce any challenge on top of those we have discussed in the case of model potentials. The FES calculated with these weights is presented in Fig.~\ref{fig:LJ-38}, where we also reported the structures of the two metastable minima corresponding to $n_6,n_8$=[14.5,13] and $n_6,n_8$=[24.5,1] respectively.

\section{Conclusions}

We introduced the iterative trajectory reweighting (ITRE) scheme, a method to evaluate the weights necessary to calculate an unbiased distribution from a biased trajectory in cases when an external, time-dependent potential is acting on the system.
ITRE does not rely on any assumption on the form of the bias potential, as it only requires being able to compute, for an arbitrary configuration of the system, the bias at a given point in time.
ITRE can be used by estimating the time-dependent biased probability distribution on a grid, but can be also implemented by simply computing averages  over configurations that had been previously encountered along the trajectory.
Thanks to the fact that it does not require the use of a regular grid, ITRE can be used even in cases where a large number of CVs are needed to bias the calculation, as we demonstrate on a challenging 6D analytical model potential.

Besides these fundamental, structural advantages, ITRE also compares favorably with existing reweighting schemes in terms of the accuracy of the reconstructed FES.
Careful tests on model potentials for which the exact FES can be computed analytically show that ITRE estimates suffer considerably less from the systematic bias that is introduced by incorporating into the average the part of the trajectory that has not yet reached a quasi-stationary distribution.
We also demonstrate that ITRE is applicable to realistic simulations by using it to reweight a trajectory of a LJ-38 cluster.

By removing some of the limitations of previous reweighting techniques in terms of the form and dimensionality of the bias potential, ITRE opens up the possibility of extracting unbiased observables from enhanced sampling calculations for which the relationship between the bias and the free energy is not straightforward.
We expect that these features will be especially useful in combination with the many, recently-proposed~\cite{awasthi2017exploring,tribello2012using,fu2018lifting},  schemes that rely on machine learning to construct high-dimensional biasing potentials.

\section{Acknowledgments}
FG, and MC were supported by the European Research Council under the European Union's Horizon 2020 research and innovation programme (Grant Agreement No. 677013-HBMAP). BC acknowledges funding from SNSF (Project P2ELP2-184408).

\bibliography{ref-mc}

\providecommand{\noopsort}[1]{}\providecommand{\singleletter}[1]{#1}%
\providecommand{\latin}[1]{#1}
\makeatletter
\providecommand{\doi}
  {\begingroup\let\do\@makeother\dospecials
  \catcode`\{=1 \catcode`\}=2 \doi@aux}
\providecommand{\doi@aux}[1]{\endgroup\texttt{#1}}
\makeatother
\providecommand*\mcitethebibliography{\thebibliography}
\csname @ifundefined\endcsname{endmcitethebibliography}
  {\let\endmcitethebibliography\endthebibliography}{}
\begin{mcitethebibliography}{37}
\providecommand*\natexlab[1]{#1}
\providecommand*\mciteSetBstSublistMode[1]{}
\providecommand*\mciteSetBstMaxWidthForm[2]{}
\providecommand*\mciteBstWouldAddEndPuncttrue
  {\def\EndOfBibitem{\unskip.}}
\providecommand*\mciteBstWouldAddEndPunctfalse
  {\let\EndOfBibitem\relax}
\providecommand*\mciteSetBstMidEndSepPunct[3]{}
\providecommand*\mciteSetBstSublistLabelBeginEnd[3]{}
\providecommand*\EndOfBibitem{}
\mciteSetBstSublistMode{f}
\mciteSetBstMaxWidthForm{subitem}{(\alph{mcitesubitemcount})}
\mciteSetBstSublistLabelBeginEnd
  {\mcitemaxwidthsubitemform\space}
  {\relax}
  {\relax}

\bibitem[Pietrucci \latin{et~al.}(2009)Pietrucci, Marinelli, Carloni, and
  Laio]{pietrucci2009substrate}
Pietrucci,~F.; Marinelli,~F.; Carloni,~P.; Laio,~A. Substrate binding mechanism
  of HIV-1 protease from explicit-solvent atomistic simulations. \emph{Journal
  of the American Chemical Society} \textbf{2009}, \emph{131},
  11811--11818\relax
\mciteBstWouldAddEndPuncttrue
\mciteSetBstMidEndSepPunct{\mcitedefaultmidpunct}
{\mcitedefaultendpunct}{\mcitedefaultseppunct}\relax
\EndOfBibitem
\bibitem[Palmer \latin{et~al.}(2014)Palmer, Martelli, Liu, Car,
  Panagiotopoulos, and Debenedetti]{palmer2014metastable}
Palmer,~J.~C.; Martelli,~F.; Liu,~Y.; Car,~R.; Panagiotopoulos,~A.~Z.;
  Debenedetti,~P.~G. Metastable liquid--liquid transition in a molecular model
  of water. \emph{Nature} \textbf{2014}, \emph{510}, 385\relax
\mciteBstWouldAddEndPuncttrue
\mciteSetBstMidEndSepPunct{\mcitedefaultmidpunct}
{\mcitedefaultendpunct}{\mcitedefaultseppunct}\relax
\EndOfBibitem
\bibitem[Khaliullin \latin{et~al.}(2011)Khaliullin, Eshet, K{\"u}hne, Behler,
  and Parrinello]{khaliullin2011nucleation}
Khaliullin,~R.~Z.; Eshet,~H.; K{\"u}hne,~T.~D.; Behler,~J.; Parrinello,~M.
  Nucleation mechanism for the direct graphite-to-diamond phase transition.
  \emph{Nature materials} \textbf{2011}, \emph{10}, 693\relax
\mciteBstWouldAddEndPuncttrue
\mciteSetBstMidEndSepPunct{\mcitedefaultmidpunct}
{\mcitedefaultendpunct}{\mcitedefaultseppunct}\relax
\EndOfBibitem
\bibitem[Angioletti-Uberti \latin{et~al.}(2010)Angioletti-Uberti, Ceriotti,
  Lee, and Finnis]{angi+10prb}
Angioletti-Uberti,~S.; Ceriotti,~M.; Lee,~P. P.~D.; Finnis,~M. W.~M.
  {Solid-liquid interface free energy through metadynamics simulations}.
  \emph{Phys. Rev. B} \textbf{2010}, \emph{81}, 125416\relax
\mciteBstWouldAddEndPuncttrue
\mciteSetBstMidEndSepPunct{\mcitedefaultmidpunct}
{\mcitedefaultendpunct}{\mcitedefaultseppunct}\relax
\EndOfBibitem
\bibitem[Fleming \latin{et~al.}(2016)Fleming, Tiwary, and
  Pfaendtner]{fleming2016new}
Fleming,~K.~L.; Tiwary,~P.; Pfaendtner,~J. New approach for investigating
  reaction dynamics and rates with ab initio calculations. \emph{The Journal of
  Physical Chemistry A} \textbf{2016}, \emph{120}, 299--305\relax
\mciteBstWouldAddEndPuncttrue
\mciteSetBstMidEndSepPunct{\mcitedefaultmidpunct}
{\mcitedefaultendpunct}{\mcitedefaultseppunct}\relax
\EndOfBibitem
\bibitem[Pietrucci and Saitta(2015)Pietrucci, and
  Saitta]{pietrucci2015formamide}
Pietrucci,~F.; Saitta,~A.~M. Formamide reaction network in gas phase and
  solution via a unified theoretical approach: Toward a reconciliation of
  different prebiotic scenarios. \emph{Proceedings of the National Academy of
  Sciences} \textbf{2015}, \emph{112}, 15030--15035\relax
\mciteBstWouldAddEndPuncttrue
\mciteSetBstMidEndSepPunct{\mcitedefaultmidpunct}
{\mcitedefaultendpunct}{\mcitedefaultseppunct}\relax
\EndOfBibitem
\bibitem[Torrie and Valleau(1977)Torrie, and Valleau]{torrie1977nonphysical}
Torrie,~G.~M.; Valleau,~J.~P. Nonphysical sampling distributions in Monte Carlo
  free-energy estimation: Umbrella sampling. \emph{Journal of Computational
  Physics} \textbf{1977}, \emph{23}, 187--199\relax
\mciteBstWouldAddEndPuncttrue
\mciteSetBstMidEndSepPunct{\mcitedefaultmidpunct}
{\mcitedefaultendpunct}{\mcitedefaultseppunct}\relax
\EndOfBibitem
\bibitem[Torrie and Valleau(1977)Torrie, and Valleau]{torrie1977monte}
Torrie,~G.; Valleau,~J. Monte Carlo study of a phase-separating liquid mixture
  by umbrella sampling. \emph{The Journal of chemical physics} \textbf{1977},
  \emph{66}, 1402--1408\relax
\mciteBstWouldAddEndPuncttrue
\mciteSetBstMidEndSepPunct{\mcitedefaultmidpunct}
{\mcitedefaultendpunct}{\mcitedefaultseppunct}\relax
\EndOfBibitem
\bibitem[Laio and Parrinello(2002)Laio, and Parrinello]{laio2002escaping}
Laio,~A.; Parrinello,~M. Escaping free-energy minima. \emph{Proceedings of the
  National Academy of Sciences} \textbf{2002}, \emph{99}, 12562--12566\relax
\mciteBstWouldAddEndPuncttrue
\mciteSetBstMidEndSepPunct{\mcitedefaultmidpunct}
{\mcitedefaultendpunct}{\mcitedefaultseppunct}\relax
\EndOfBibitem
\bibitem[Barducci \latin{et~al.}(2008)Barducci, Bussi, and
  Parrinello]{barducci2008well}
Barducci,~A.; Bussi,~G.; Parrinello,~M. Well-tempered metadynamics: a smoothly
  converging and tunable free-energy method. \emph{Physical review letters}
  \textbf{2008}, \emph{100}, 020603\relax
\mciteBstWouldAddEndPuncttrue
\mciteSetBstMidEndSepPunct{\mcitedefaultmidpunct}
{\mcitedefaultendpunct}{\mcitedefaultseppunct}\relax
\EndOfBibitem
\bibitem[Piana and Laio(2007)Piana, and Laio]{piana2007bias}
Piana,~S.; Laio,~A. A bias-exchange approach to protein folding. \emph{The
  journal of physical chemistry B} \textbf{2007}, \emph{111}, 4553--4559\relax
\mciteBstWouldAddEndPuncttrue
\mciteSetBstMidEndSepPunct{\mcitedefaultmidpunct}
{\mcitedefaultendpunct}{\mcitedefaultseppunct}\relax
\EndOfBibitem
\bibitem[Pfaendtner and Bonomi(2015)Pfaendtner, and
  Bonomi]{pfaendtner2015efficient}
Pfaendtner,~J.; Bonomi,~M. Efficient sampling of high-dimensional free-energy
  landscapes with parallel bias metadynamics. \emph{Journal of chemical theory
  and computation} \textbf{2015}, \emph{11}, 5062--5067\relax
\mciteBstWouldAddEndPuncttrue
\mciteSetBstMidEndSepPunct{\mcitedefaultmidpunct}
{\mcitedefaultendpunct}{\mcitedefaultseppunct}\relax
\EndOfBibitem
\bibitem[Valsson and Parrinello(2014)Valsson, and
  Parrinello]{valsson2014variational}
Valsson,~O.; Parrinello,~M. Variational approach to enhanced sampling and free
  energy calculations. \emph{Physical review letters} \textbf{2014},
  \emph{113}, 090601\relax
\mciteBstWouldAddEndPuncttrue
\mciteSetBstMidEndSepPunct{\mcitedefaultmidpunct}
{\mcitedefaultendpunct}{\mcitedefaultseppunct}\relax
\EndOfBibitem
\bibitem[Whitmer \latin{et~al.}(2014)Whitmer, Chiu, Joshi, and
  De~Pablo]{whitmer2014basis}
Whitmer,~J.~K.; Chiu,~C.-c.; Joshi,~A.~A.; De~Pablo,~J.~J. Basis function
  sampling: A new paradigm for material property computation. \emph{Physical
  review letters} \textbf{2014}, \emph{113}, 190602\relax
\mciteBstWouldAddEndPuncttrue
\mciteSetBstMidEndSepPunct{\mcitedefaultmidpunct}
{\mcitedefaultendpunct}{\mcitedefaultseppunct}\relax
\EndOfBibitem
\bibitem[Sidky and Whitmer(2018)Sidky, and Whitmer]{sidky2018learning}
Sidky,~H.; Whitmer,~J.~K. Learning free energy landscapes using artificial
  neural networks. \emph{The Journal of chemical physics} \textbf{2018},
  \emph{148}, 104111\relax
\mciteBstWouldAddEndPuncttrue
\mciteSetBstMidEndSepPunct{\mcitedefaultmidpunct}
{\mcitedefaultendpunct}{\mcitedefaultseppunct}\relax
\EndOfBibitem
\bibitem[Tiwary and Parrinello(2014)Tiwary, and Parrinello]{tiwary2014time}
Tiwary,~P.; Parrinello,~M. A time-independent free energy estimator for
  metadynamics. \emph{The Journal of Physical Chemistry B} \textbf{2014},
  \emph{119}, 736--742\relax
\mciteBstWouldAddEndPuncttrue
\mciteSetBstMidEndSepPunct{\mcitedefaultmidpunct}
{\mcitedefaultendpunct}{\mcitedefaultseppunct}\relax
\EndOfBibitem
\bibitem[Bonomi \latin{et~al.}(2009)Bonomi, Barducci, and
  Parrinello]{bonomi2009reconstructing}
Bonomi,~M.; Barducci,~A.; Parrinello,~M. Reconstructing the equilibrium
  Boltzmann distribution from well-tempered metadynamics. \emph{Journal of
  computational chemistry} \textbf{2009}, \emph{30}, 1615--1621\relax
\mciteBstWouldAddEndPuncttrue
\mciteSetBstMidEndSepPunct{\mcitedefaultmidpunct}
{\mcitedefaultendpunct}{\mcitedefaultseppunct}\relax
\EndOfBibitem
\bibitem[Tiana(2008)]{tiana2008estimation}
Tiana,~G. Estimation of microscopic averages from metadynamics. \emph{The
  European Physical Journal B} \textbf{2008}, \emph{63}, 235--238\relax
\mciteBstWouldAddEndPuncttrue
\mciteSetBstMidEndSepPunct{\mcitedefaultmidpunct}
{\mcitedefaultendpunct}{\mcitedefaultseppunct}\relax
\EndOfBibitem
\bibitem[Awasthi and Nair(2017)Awasthi, and Nair]{awasthi2017exploring}
Awasthi,~S.; Nair,~N.~N. Exploring high dimensional free energy landscapes:
  Temperature accelerated sliced sampling. \emph{The Journal of Chemical
  Physics} \textbf{2017}, \emph{146}, 094108\relax
\mciteBstWouldAddEndPuncttrue
\mciteSetBstMidEndSepPunct{\mcitedefaultmidpunct}
{\mcitedefaultendpunct}{\mcitedefaultseppunct}\relax
\EndOfBibitem
\bibitem[Tribello \latin{et~al.}(2012)Tribello, Ceriotti, and
  Parrinello]{tribello2012using}
Tribello,~G.~A.; Ceriotti,~M.; Parrinello,~M. Using sketch-map coordinates to
  analyze and bias molecular dynamics simulations. \emph{Proceedings of the
  National Academy of Sciences} \textbf{2012}, \emph{109}, 5196--5201\relax
\mciteBstWouldAddEndPuncttrue
\mciteSetBstMidEndSepPunct{\mcitedefaultmidpunct}
{\mcitedefaultendpunct}{\mcitedefaultseppunct}\relax
\EndOfBibitem
\bibitem[Fu and Pfaendtner(2018)Fu, and Pfaendtner]{fu2018lifting}
Fu,~C.~D.; Pfaendtner,~J. Lifting the curse of dimensionality on enhanced
  sampling of reaction networks with parallel bias metadynamics. \emph{Journal
  of chemical theory and computation} \textbf{2018}, \emph{14},
  2516--2525\relax
\mciteBstWouldAddEndPuncttrue
\mciteSetBstMidEndSepPunct{\mcitedefaultmidpunct}
{\mcitedefaultendpunct}{\mcitedefaultseppunct}\relax
\EndOfBibitem
\bibitem[Tribello \latin{et~al.}(2014)Tribello, Bonomi, Branduardi, Camilloni,
  and Bussi]{tribello2014plumed}
Tribello,~G.~A.; Bonomi,~M.; Branduardi,~D.; Camilloni,~C.; Bussi,~G. PLUMED 2:
  New feathers for an old bird. \emph{Computer Physics Communications}
  \textbf{2014}, \emph{185}, 604--613\relax
\mciteBstWouldAddEndPuncttrue
\mciteSetBstMidEndSepPunct{\mcitedefaultmidpunct}
{\mcitedefaultendpunct}{\mcitedefaultseppunct}\relax
\EndOfBibitem
\bibitem[Valsson \latin{et~al.}(2016)Valsson, Tiwary, and
  Parrinello]{valsson2016enhancing}
Valsson,~O.; Tiwary,~P.; Parrinello,~M. Enhancing important fluctuations: Rare
  events and metadynamics from a conceptual viewpoint. \emph{Annual review of
  physical chemistry} \textbf{2016}, \emph{67}, 159--184\relax
\mciteBstWouldAddEndPuncttrue
\mciteSetBstMidEndSepPunct{\mcitedefaultmidpunct}
{\mcitedefaultendpunct}{\mcitedefaultseppunct}\relax
\EndOfBibitem
\bibitem[Chen \latin{et~al.}(2019)Chen, Sidky, and Ferguson]{chen2019nonlinear}
Chen,~W.; Sidky,~H.; Ferguson,~A.~L. Nonlinear discovery of slow molecular
  modes using state-free reversible VAMPnets. \emph{The Journal of chemical
  physics} \textbf{2019}, \emph{150}, 214114\relax
\mciteBstWouldAddEndPuncttrue
\mciteSetBstMidEndSepPunct{\mcitedefaultmidpunct}
{\mcitedefaultendpunct}{\mcitedefaultseppunct}\relax
\EndOfBibitem
\bibitem[Mendels \latin{et~al.}(2018)Mendels, Piccini, and
  Parrinello]{mendels2018collective}
Mendels,~D.; Piccini,~G.; Parrinello,~M. Collective variables from local
  fluctuations. \emph{The journal of physical chemistry letters} \textbf{2018},
  \emph{9}, 2776--2781\relax
\mciteBstWouldAddEndPuncttrue
\mciteSetBstMidEndSepPunct{\mcitedefaultmidpunct}
{\mcitedefaultendpunct}{\mcitedefaultseppunct}\relax
\EndOfBibitem
\bibitem[Chen \latin{et~al.}(2018)Chen, Tan, and Ferguson]{chen2018collective}
Chen,~W.; Tan,~A.~R.; Ferguson,~A.~L. Collective variable discovery and
  enhanced sampling using autoencoders: Innovations in network architecture and
  error function design. \emph{The Journal of chemical physics} \textbf{2018},
  \emph{149}, 072312\relax
\mciteBstWouldAddEndPuncttrue
\mciteSetBstMidEndSepPunct{\mcitedefaultmidpunct}
{\mcitedefaultendpunct}{\mcitedefaultseppunct}\relax
\EndOfBibitem
\bibitem[Gimondi \latin{et~al.}(2018)Gimondi, Tribello, and
  Salvalaglio]{gimondi2018building}
Gimondi,~I.; Tribello,~G.~A.; Salvalaglio,~M. Building maps in collective
  variable space. \emph{The Journal of chemical physics} \textbf{2018},
  \emph{149}, 104104\relax
\mciteBstWouldAddEndPuncttrue
\mciteSetBstMidEndSepPunct{\mcitedefaultmidpunct}
{\mcitedefaultendpunct}{\mcitedefaultseppunct}\relax
\EndOfBibitem
\bibitem[Marinova and Salvalaglio(2019)Marinova, and
  Salvalaglio]{marinova2019time}
Marinova,~V.; Salvalaglio,~M. Time-independent free energies from metadynamics
  via Mean Force Integration. \emph{arXiv preprint arXiv:1907.08472}
  \textbf{2019}, \relax
\mciteBstWouldAddEndPunctfalse
\mciteSetBstMidEndSepPunct{\mcitedefaultmidpunct}
{}{\mcitedefaultseppunct}\relax
\EndOfBibitem
\bibitem[Bonomi \latin{et~al.}(2009)Bonomi, Branduardi, Bussi, Camilloni,
  Provasi, Raiteri, Donadio, Marinelli, Pietrucci, Broglia, and
  Parrinello]{PLUMED}
Bonomi,~M.; Branduardi,~D.; Bussi,~G.; Camilloni,~C.; Provasi,~D.; Raiteri,~P.;
  Donadio,~D.; Marinelli,~F.; Pietrucci,~F.; Broglia,~R.~A. \latin{et~al.}
  {PLUMED: A portable plugin for free-energy calculations with molecular
  dynamics}. \emph{Comp. Phys. Comm.} \textbf{2009}, \emph{180},
  1961--1972\relax
\mciteBstWouldAddEndPuncttrue
\mciteSetBstMidEndSepPunct{\mcitedefaultmidpunct}
{\mcitedefaultendpunct}{\mcitedefaultseppunct}\relax
\EndOfBibitem
\bibitem[Roux(1995)]{roux1995calculation}
Roux,~B. The calculation of the potential of mean force using computer
  simulations. \emph{Computer physics communications} \textbf{1995}, \emph{91},
  275--282\relax
\mciteBstWouldAddEndPuncttrue
\mciteSetBstMidEndSepPunct{\mcitedefaultmidpunct}
{\mcitedefaultendpunct}{\mcitedefaultseppunct}\relax
\EndOfBibitem
\bibitem[Kumar \latin{et~al.}(1992)Kumar, Rosenberg, Bouzida, Swendsen, and
  Kollman]{kumar1992weighted}
Kumar,~S.; Rosenberg,~J.~M.; Bouzida,~D.; Swendsen,~R.~H.; Kollman,~P.~A. The
  weighted histogram analysis method for free-energy calculations on
  biomolecules. I. The method. \emph{Journal of computational chemistry}
  \textbf{1992}, \emph{13}, 1011--1021\relax
\mciteBstWouldAddEndPuncttrue
\mciteSetBstMidEndSepPunct{\mcitedefaultmidpunct}
{\mcitedefaultendpunct}{\mcitedefaultseppunct}\relax
\EndOfBibitem
\bibitem[Bonomi \latin{et~al.}(2019)Bonomi, Bussi, Camilloni, Tribello,
  Ban{\'a}{\v{s}}, Barducci, Bernetti, Bolhuis, Bottaro, Branduardi,
  \latin{et~al.} others]{bonomi2019promoting}
Bonomi,~M.; Bussi,~G.; Camilloni,~C.; Tribello,~G.~A.; Ban{\'a}{\v{s}},~P.;
  Barducci,~A.; Bernetti,~M.; Bolhuis,~P.~G.; Bottaro,~S.; Branduardi,~D.
  \latin{et~al.}  Promoting transparency and reproducibility in enhanced
  molecular simulations. \emph{Nature methods} \textbf{2019}, \emph{16},
  670--673\relax
\mciteBstWouldAddEndPuncttrue
\mciteSetBstMidEndSepPunct{\mcitedefaultmidpunct}
{\mcitedefaultendpunct}{\mcitedefaultseppunct}\relax
\EndOfBibitem
\bibitem[Plimpton(1995)]{plim95jcp}
Plimpton,~S. {Fast Parallel Algorithms for Short-Range Molecular Dynamics}.
  \emph{J. Comp. Phys.} \textbf{1995}, \emph{117}, 1--19\relax
\mciteBstWouldAddEndPuncttrue
\mciteSetBstMidEndSepPunct{\mcitedefaultmidpunct}
{\mcitedefaultendpunct}{\mcitedefaultseppunct}\relax
\EndOfBibitem
\bibitem[Ceriotti \latin{et~al.}(2011)Ceriotti, Brain, Riordan, and
  Manolopoulos]{ceri+12prsa}
Ceriotti,~M.; Brain,~G. A.~R.; Riordan,~O.; Manolopoulos,~D.~E. {The
  inefficiency of re-weighted sampling and the curse of system size in high
  order path integration}. \emph{Proceedings of the Royal Society A:
  Mathematical, Physical and Engineering Sciences} \textbf{2011}, \emph{468},
  2--17\relax
\mciteBstWouldAddEndPuncttrue
\mciteSetBstMidEndSepPunct{\mcitedefaultmidpunct}
{\mcitedefaultendpunct}{\mcitedefaultseppunct}\relax
\EndOfBibitem
\bibitem[Kullback and Leibler(1951)Kullback, and
  Leibler]{kullback1951information}
Kullback,~S.; Leibler,~R.~A. On information and sufficiency. \emph{The annals
  of mathematical statistics} \textbf{1951}, \emph{22}, 79--86\relax
\mciteBstWouldAddEndPuncttrue
\mciteSetBstMidEndSepPunct{\mcitedefaultmidpunct}
{\mcitedefaultendpunct}{\mcitedefaultseppunct}\relax
\EndOfBibitem
\bibitem[Branduardi \latin{et~al.}(2012)Branduardi, Bussi, and
  Parrinello]{branduardi2012metadynamics}
Branduardi,~D.; Bussi,~G.; Parrinello,~M. Metadynamics with adaptive Gaussians.
  \emph{Journal of chemical theory and computation} \textbf{2012}, \emph{8},
  2247--2254\relax
\mciteBstWouldAddEndPuncttrue
\mciteSetBstMidEndSepPunct{\mcitedefaultmidpunct}
{\mcitedefaultendpunct}{\mcitedefaultseppunct}\relax
\EndOfBibitem
\end{mcitethebibliography}

\end{document}


\maketitle

\section{\label{sec:potential}Analytic Potential}

To illustrate the efficiency of equation \ref{eq:time_c_t} we performed several Well-Tempered Metadynamics calculations for a Langevin particle in an analytic potential with a different number of dimensions.

We designed an analytic potential energy surface $U(\mathbf{x})$ composed by three terms: minima shaped as narrow channels $C^i(\mathbf{x})$, points connecting different minima in a smooth way $P^j(\mathbf{x})$, and Gaussian functions $G^k(\mathbf{x})$ which represent barriers. While the point and channel contribution are attractive, the Gaussian contributions are purely repulsive, so $P^j(\mathbf{x})$ and $C^i(\mathbf{x})$ are used to shape the minima of the potential, while $G^k(\mathbf{x})$ are used to tune the potential energy barriers. The functional form of the points is straightforward:

\begin{equation}
P^j(\mathbf{x}) = 1 + \sum_a^D \frac{(x_a - p^j_a)^2}{{\sigma^j}^2}
\label{eq:point}
\end{equation}

where $D$ is the number of dimensions, and $a$ the index of the coordinate, while $\mathbf{p}$ and $\sigma^j$ are respectively the center of the points and the its "width". This potential is one near the center and increases quadratically as the distance from the center increases. The barriers are simple Gaussian functions:

\begin{equation}
G^k(\mathbf{x}) = h^k e^{-\frac{(\mathbf{x} - \mathbf{\mu}^k)^2}{2 {\sigma^k}^2}}
\label{eq:gaussian}
\end{equation}

with $\mu^k$ the center and $\sigma^k$ the width of the $k$th Gaussian. The form for $C^i(\mathbf{x})$ is slightly more complex:

\begin{equation}
C^i(\mathbf{x}) = \sum_a^D \left (\frac{(x_a-c^i_a)\mathbf{r}^i_a}{\sigma_a^i} \right )^{n^i_a}
\label{eq:minimun}
\end{equation}

The center of the minimum is identified by the vector $\mathbf{c}^i$, while $\mathbf{\sigma}^i_a$, $\mathbf{r}^i$, and $n_a^i$ are respectively a length contraction, a rotation matrix, and a real number, that can be used to orient and shape the minimum in space. As for the points, $C^i(\mathbf{x})$ has a minimum in the center and increases if we increase our distance from it.

While it can be tempting to simply sum the three contributions together, this would not be optimal. Far from the point and minimum center, the potential quickly increases to infinity, making negligible the contribution of other terms. The energy value of the minima and barriers of the potential depend on the position of each other, making it difficult to use. To avoid this problem and have a better behaving potential we combine them as

\begin{eqnarray}
U(\mathbf{x}) = & \alpha & f^{-1} \left (\sum_i f(C^i(\mathbf{x})) + \sum_j f(P^j(\mathbf{x})) \right ) \\ \nonumber
+ & \alpha & \sum_k G^k(\mathbf{x})
\label{eq:U_tot}
\end{eqnarray}

where $\alpha$ is a scaling parameter. The effect of the application of consequently apply $f$ and $f^{-1}$, if $f$ is chosen properly, is to avoid the potential to increase uncontrollably and place all the channels at exactly $U(\mathbf{c}^i)=0$. In this work, we used $f=1/x$ and $f^{-1}=1/f$. An example of the potential for a $D=2$ system with three minima is reported in figure \sref{SI-fig:pot_example}:

\begin{figure}[ht]
    \centering
    \includegraphics[width=\textwidth]{figures/example_potential.pdf}
    \caption{Example of the potential $U(\mathbf{x})$ in 2D composed of three minima at $\mathbf{c}= [0.0,0.0]; [0.0,-1.0]; [1.0,0.0]$.}
    \label{SI-fig:pot_example}
\end{figure}

\section{Evaluation of $c_I(t)$ with the equilibrium P(\bs)}

In the manuscript in section \ref{sec:method}
we illustrate that it is necessary to used $P(\bs,t)$ rather than the equilibrium distribution $P(\bs)$ to evaluate $c(t)$ correctly. Here we illustrate what happens if the equilibrium distribution $P(\bs)$ is used. Let's start by noticing that at the beginning of the calculation the potential $V(s(t),t=0)$ is zero, which means that $c_I(t=0)$ is also zero, as the numerator and denominator in Eqn.~\eqref{eq:probab_omega} are equal. As we explore the first minimum $V(s(t),t)$ increases, the nominator of Eqn.~\eqref{eq:probab_omega} decreases, meaning that $c_I(t)$ increases. However, as long as the system is trapped in the minimum, the contribution of this minimum in Eqn.~\eqref{eq:probab_omega} is quickly set to zero as $V(\bs(t),t)$ will continuously increase. This means that $c_I(t)$ plateaus, and does not change until we escape and explore a new minimum. The same situation is eventually repeated, and only at the end of the calculation when all the minima have been visited and $V(\bs(t),t=T)$ is large in the entire CVs space, $c_I(t)$ will tend to a maximum, as the numerator of equation \ref{eq:probab_omega} will tend to zero. Of course, this is true if $P(\bs)$ is evaluated on the entire CVs space, or in other words, the full trajectory is used to evaluate $P(\bs)$. If we evaluate $P(\bs)$ at an earlier time, corresponding to only part of the CVs space, $c_I(t)$ nonetheless saturates, but at a different value. We reported different examples of $c_I(t)$ in which the probability distribution was calculated only using the subset of the trajectory that runs up to $T$. For $T=1800$, we only have statistics from the first minimum, and $c_I(t)$ does not saturate but grows monotonically. For larger $T$, we start to collect statistics from two and three minima, and $c_I(t)$ saturates at different values while the system does not explore the entire CVs space. Notice that for $T=15000$ and $T=30000$, $c_I(t)$ changes drastically at any time. The reason is that after this amount of time, all the large probability states have been explored and only low probable, high energy states remain.

\begin{figure}
    \centering
    \includegraphics[width=\textwidth]{figures/full_c_t.pdf}
    \caption{Different calculation of $c_I(t)$ where the upper limit of the integration $T$ is shifted, using the same potential $V(\bs(t),t)$. As explained in the text, $c_I(t)$ always reaches a maximum for $t=T$. However, how $c_I(t)$ increases depends on how much of the phase space has been explored up to  $T$. So for example, when $T=1800$ we are trapped in the first minimum and $c_I(t)$ quickly rise and reaches a maximum. However, for $T=3600$, $c_I(t)$ saturate for $t<2000$ and jumps when we move from the first minimum to the second. Notice also, that the value for which $c_I(t)$ plateau when trapped inside minima changes if we change $T$.}
    \label{fig:full_c_t}
\end{figure}

If $c_I(t)$ is evaluated in such a way, it can provide a useful diagnostic on whether or not our calculations have surpassed the transient. Times in which $t$ where $c_I(t)$ exhibit jumps and plateaus are still part of the transient.

\section{Convergence of c(t)}

In Sec.~\ref{subsec:convergence} of the manuscript we illustrate how $c_I(t)$ is robust with respect to the number of iteration and the stride used to calculate it. Here we also reported the absolute error evaluated with respect to the analytic FES for the $D=2$ system

\begin{figure}
    \centering
    \includegraphics[width=0.78\textwidth]{figures/convergence.pdf}
    \caption{Absolute error $F(\bs)-\tilde{F}(\bs)$ between the ITRE evaluated FES $F(\bs)$ and the analytic $\tilde{F}(\bs)$ with respect to $I$ (upper panel) and $\tau$ (lower panel). As can be seen, the error converges in both cases: in the former case, we obtained convergence after the 3rd iteration. In the latter, to converge the error is necessary to use a stride of maximum 100 kernel deposition. }
    \label{fig:convergence_error}
\end{figure}

In Sec.~\ref{subsec:convergence} we also explain that the weights calculated with ITRE can increase substantially if the WTM calculation is performed with Gaussian kernels that are too large. In Fig.~\sref{fig:weights_large} we illustrate how the ITRE and $c_s(t)$ weights compare if the Gaussian used are of a width size comparable with the width of the basins characterizing the FES.

\begin{figure}
\centering
\includegraphics[width=0.65\textwidth]{figures/c_t_large.pdf}
    \caption{Comparison between $c_I(t)$ and $c_s(t)$ for a Gaussian with large widths. Although the same consideration that we draw for Fig.~\ref{fig:weights} holds, as $c_I(t)$ has lower weights at the beginning of the trajectory that at the end, they are larger due to the larger overlaps between the Gaussian kernels. }
    \label{fig:weights_large}
\end{figure}

\section{2D systems}



In Sec.~\ref{sec:results} of the manuscript we discuss the results for the $D=3$ system only. Here we also present the results for the $D=2$ case, for which both methods converge the probability function to an accurate result, as can be seen in \ref{fig:kl_fine_2d}. The FES is almost indistinguishable from the analytic one, apart from a small difference near the barrier in the mono-dimensional one.

\begin{figure}
    \includegraphics[width=0.9\textwidth]{figures/KL_fine_comparison.pdf}
    \caption{$D_{KL}$ measure for the mono-dimensional $P(\bs)$ (top and lower panel) for the $D=2$ system. In all cases the KL divergence decreases with the proceeding of time, and both reweigthing methods provide almost the same result in the long time limit.}
    \label{fig:kl_fine_2d}
\end{figure}

\begin{figure*}
    \includegraphics[width=\textwidth]{figures/FES_fine_comparison.pdf}
    \caption{FES for the $D=2$ system calculated as  a function of $x$ and $y$ respectively (left and right panels). The analytic FES is reported in black, with no error bars, while in blue and yellow the $c_I(t)$ and $c_s(t)$ results.}
    \label{fig:fes2_d}
\end{figure*}

We also present the data obtained from four independent calculations performed with diffusive adaptive hills, for which we also calculated the $D_{KL}$ divergence as well as the FES.

\begin{figure}
    \includegraphics[width=0.9\textwidth]{figures/KL_adaptive.pdf}
    \caption{$D_{KL}$ measure for the mono-dimensional $P(s)$ (top and lower panels) for the $D=2$ system with the diffusive adaptive Gaussian scheme.}
    \label{fig:kl_adaptive}
\end{figure}

\begin{figure*}
    \includegraphics[width=\textwidth]{figures/FES_adaptive.pdf}
    \caption{FES for the $D=2$ system calculated as  a function of $x$ and $y$ respectively (left and right panels) with the diffusive adaptive gaussians scheme. The analytic FES is reported in black, with no error bars, while in blue and yellow the $c_I(t)$ and $c_s(t)$ results.}
    \label{fig:fes2_adaptive}
\end{figure*}

The potential energy surface used to perform these simulations was constructed using the following parameters for the equations introduce in section \sref{sec:potential}.

The minima were constructed with the following parameters

\begin{eqnarray}
&\mathbf{c}_1& = [0,-1] \quad \boldsymbol{\sigma}_1=[1.0,0.2]
\quad \mathbf{n}_1=[8,2]
\\ \nonumber
&\mathbf{c}_2& = [1,0] \quad \boldsymbol{\sigma}_2=[0.2,1.0]
\quad \mathbf{n}_2=[2,8]
\\\nonumber
&\mathbf{c}_3& = [0,0] \quad \boldsymbol{\sigma}_3=[\sqrt{2},0.2]
\quad \mathbf{n}_3=[8,2]
\end{eqnarray}

The first and second minima were not rotated, while the third was rotated using the following rotation matrix

\begin{equation}
\mathbf{R}_1 =
\begin{bmatrix}
\frac{\sqrt{2}}{2} & \frac{\sqrt{2}}{2} \\
-\frac{\sqrt{2}}{2} & \frac{\sqrt{2}}{2}
\end{bmatrix}
\end{equation}

To connect the minima, we used points defined with the following parameters

\begin{eqnarray}
\mathbf{p}_1 &=& [1,-1] \quad \boldsymbol{\sigma}_1=[0.2,0.2]\\\nonumber
\mathbf{p}_2 &=& [1,1] \quad \boldsymbol{\sigma}_2=[0.2,0.2]\\\nonumber
\mathbf{p}_3 &=& [-1,1] \quad \boldsymbol{\sigma}_3=[0.2,0.2]
\end{eqnarray}

To introduce barriers between the minima, gaussian functions were introduced on top of points with the same parameters as the points and height $h=1$.

The total potential is then scaled with a $\alpha=30$.

\section{System $D=3$}

In Sec.~\ref{sec:results} of the manuscript we illustrated how for the $D=3$ system, the method is capable of evaluating the bi-dimensional probability density function and FES and achieve convergence faster than Valsson \textit{et al.} implementation. Here we also reported the $D_{KL}$ divergence and FES for the mono-dimensional case, showing how for x and y ITRE is faster than $c_s(t)$ in evaluating the probability density function. For z, both methods recover the correct probability density function very rapidly.

\begin{figure}[ht]
    \centering
    \includegraphics[width=0.9\textwidth]{figures/KL_fine_comparison_1d.pdf}
    \caption{Kullback-Leibler divergence for P(x), P(y), and P(z) for ITRE and $c_s(t)$ reported in the top, central and lower panel respectively. While for z the two methods converge quite rapidly, for x and y ITRE achieve a better convergence at a faster rate than $c_s(t)$.}
    \label{fig:KL_3d_mono}
\end{figure}

In addition, we illustrate in Fig.~\sref{fig:FES_3d_mono} the mono-dimensional FES obtained as a function of $x$, $y$ and $z$ for both ITRE and $c_s(t)$, while in figure
Fig.~\sref{fig:FES_3d_bi} we compare the bi-dimensional FES as a function of $x,y$, $x,z$ and $y,z$ which are also reported in Fig.~\ref{fig:3d_system}.

\begin{figure}[ht]
    \centering
    \includegraphics[width=\textwidth]{figures/FES_fine_1d.pdf}
    \caption{FES calculated with the Iterative approach (blue line) and with the Valsson \textit{et al.} method (orange line). The analytic FES is reported in solid black. The discrepancy visible in the left and right minima is due to the systematic error that we introduced by generating our starting configuration always from the same point. This error disappear in the long time limit, as the estimate converge to the correct FES, but the large weights of early microstates (which is always the first minimum) poison the evaluated P(s) so that it requires a long time to converge.}
    \label{fig:FES_3d_mono}
\end{figure}

\begin{figure}[ht]
    \centering
    \includegraphics[width=\textwidth]{figures/FES_fine_2d.pdf}
    \caption{FES calculated with the Iterative approach (black line) and with the Valsson \textit{et al.} method (red line). The analytic FES is reported in filled contours.}
    \label{fig:FES_3d_bi}
\end{figure}

The potential energy surface for the $D=3$ system was composed using the following parameters:

\begin{eqnarray}
&\mathbf{c}_1& = [0,-1,-1] \quad \boldsymbol{\sigma}_1=[1.0,0.1,0.1] \quad \mathbf{n}_1=[8,2,2] \\ \nonumber
&\mathbf{c}_2& = [-1,-1,0] \quad \boldsymbol{\sigma}_2=[0.1,0.1,1.0] \quad \mathbf{n}_2=[2,2,8] \\\nonumber
&\mathbf{c}_3& = [0,-1,1] \quad \boldsymbol{\sigma}_3=[1.0,0.1,0.1] \quad \mathbf{n}_3=[8,2,2] \\\nonumber
&\mathbf{c}_4& = [1,0,0] \quad \boldsymbol{\sigma}_4=[0.1,\sqrt{2},0.1] \quad \mathbf{n}_4=[2,8,2] \\\nonumber
&\mathbf{c}_5& = [0,1,0] \quad \boldsymbol{\sigma}_5=[\sqrt{2},0.1,0.1] \quad \mathbf{n}_5=[8,2,2] \\\nonumber
&\mathbf{c}_6& = [0,0,0] \quad \boldsymbol{\sigma}_6=[0.1,\sqrt{3},0.1] \quad \mathbf{n}_6=[2,8,2]
\end{eqnarray}

While the first three minima were not rotated, the other were rotated according to the following rotation matrices:

\begin{eqnarray}
& \mathbf{R}_1 =
\begin{bmatrix}
1 & 0 & 0 \\
0 & \frac{\sqrt{2}}{2} & \frac{\sqrt{2}}{2} \\
0 & -\frac{\sqrt{2}}{2} & \frac{\sqrt{2}}{2}
\end{bmatrix}
\\ \nonumber
& \mathbf{R}_2 =
\begin{bmatrix}
\frac{\sqrt{2}}{2} & 0 & \frac{\sqrt{2}}{2} \\
0 & 1 & 0 \\
-\frac{\sqrt{2}}{2} & 0 & \frac{\sqrt{2}}{2}
\end{bmatrix}
\\ \nonumber
& \mathbf{R}_3 =
\begin{bmatrix}
-\frac{\sqrt{2}}{2} & -\frac{\sqrt{2}}{2} &  0.0 \\
 0.5       & -0.5       & -\frac{\sqrt{2}}{2} \\
 0.5       & -0.5       &  \frac{\sqrt{2}}{2}
\end{bmatrix}
\end{eqnarray}

To connect the minima, we used points defined with the following parameters

\begin{eqnarray}
\mathbf{p}_1 &=& [-1,-1,-1] \quad \boldsymbol{\sigma}_1=[0.2,0.2,0.2]\\\nonumber
\mathbf{p}_2 &=& [-1,-1,1] \quad \boldsymbol{\sigma}_2=[0.2,0.2,0.2]\\\nonumber
\mathbf{p}_3 &=& [1,-1,1] \quad \boldsymbol{\sigma}_3=[0.2,0.2,0.2]\\\nonumber
\mathbf{p}_4 &=& [1,1,-1] \quad \boldsymbol{\sigma}_4=[0.2,0.2,0.2]\\\nonumber
\mathbf{p}_5 &=& [-1,1,1] \quad \boldsymbol{\sigma}_5=[0.2,0.2,0.2]\\\nonumber
\mathbf{p}_6 &=& [1,-1,-1] \quad \boldsymbol{\sigma}_6=[0.2,0.2,0.2]
\end{eqnarray}

To introduce barriers between the minima, gaussian functions were introduced on top of points with the same parameters as the points and height $h=1$.

The total potential is then scaled with a $\alpha=30$.

\section{\label{SI-sec:6d}System $D=6$}

In Sec.~\ref{sec:results} of the manuscript we illustrated how the FES for a system with $D=6$ characterized by three minima can be sampled with adaptive hills. Here we show that the same FES can be sampled with the regular Gaussian kernels. We illustrate in Fig.~ \sref{fig:6d_weigths} the $c_I(t)$ and the corresponding weights obtained for the four WTM calculations, while we have reported in Fig.~\sref{fig:fes_normal_6d_mono} and \sref{fig:fes_normal_6d_bi} the mono-dimensional and bi-dimensional FES, as a function of the $d3$ $d5$ and $d6$ collective variables.

\begin{figure}
    \includegraphics[width=\textwidth]{figures/6D_weights.pdf}
    \caption{The behavior of $c_I(t)$ and the corresponding weights in the four calculations with regular Gaussian kernels in the $D=6$ system.}
    \label{fig:6d_weigths}
\end{figure}

\begin{figure}
    \includegraphics[width=\textwidth]{figures/FES_normal_6d.pdf}
    \caption{FES for the $D=6$ system calculated as  a function of $d6$, $d5$ and $d3$. The analytic FES is reported in black, while we reported the $c_I(t)$ reconstructed FES in blue.}
    \label{fig:fes_normal_6d_mono}
\end{figure}

\begin{figure}
    \includegraphics[width=\textwidth]{figures/FES_6_d_2d.pdf}
    \caption{FES for the $D=6$ system calculated as  a function of $d5,d6$, and $d3,d5$. The analytic FES is reported with filled colored contours, while in black we reported the $c_I(t)$ reconstructed FES.}
    \label{fig:fes_normal_6d_bi}
\end{figure}

\begin{figure}
    \includegraphics[width=\textwidth]{figures/FES_6_d_2d_difference.pdf}
    \caption{Differences between the analytic and reconstructed FES as a function of the $d5,d6$, and $d3,d5$ for the cases with regular hills.}
    \label{fig:fes_normal_6d_bi_diff}
\end{figure}

\begin{figure}
    \includegraphics[width=\textwidth]{figures/FES_6_d_adaptive_difference.pdf}
    \caption{Differences between the analytic and reconstructed FES as a function of the $d5,d6$, and $d3,d5$ calculated with the diffusive adaptive hills.}
    \label{fig:fes_normal_6d_bi_adap_diff}
\end{figure}

\begin{figure}
    \includegraphics[width=\textwidth]{figures/6D_weights_adaptive.pdf}
    \caption{Behavior of $c_I(t)$ and the corresponding weights in the four calculations with adaptive Gaussian kernels in the $D=6$ system.}
    \label{fig:fes_adaptive_6d_bi_weights}
\end{figure}

The potential energy surface for the $D=6$ system was composed using the following parameters:

\begin{eqnarray}
&\mathbf{c}_1& = [1.0,1.0,1.0,0.0,0.0,0.0] \\ \nonumber
&\mathbf{c}_2& = [0.0,0.0,0.0,-1.0,-1.0,-1.0]\\\nonumber
&\mathbf{c}_3& = [0.0,0.0,0.0,0.0,0.0,0.0]
\end{eqnarray}

The same $\boldsymbol{\sigma}_1=[1.0,0.2,0.2,0.2,0.2,0.2]$ and  $\mathbf{n}_1=[8,2,2,2,2,2]$ were used for all the minima. The minima were rotated according to the following rotation matrices

\begin{eqnarray}
& \mathbf{R}_1 =
\begin{bmatrix}
                    0.0 &
                    0.0&
                    0.0&
                    -0.57735&
                    -0.57735&
                    -0.57735
                \\
                    0.0&
                    1.0&
                    0.0&
                    0.0&
                    0.0&
                    0.0
                \\
                    0.0&
                    0.0&
                    1.0&
                    0.0&
                    0.0&
                    0.0
                \\
                    0.57735
                    0.0&
                    0.0&
                    0.66666&
                    -0.33333&
                    -0.33333&
                \\
                    0.57735&
                    0.0&
                    0.0&
                    -0.33333&
                    0.66666&
                    -0.33333&
                \\
                    0.57735&
                    0.0&
                    0.0&
                    -0.33333&
                    -0.33333&
                    0.66666
\end{bmatrix}
\\ \nonumber
& \mathbf{R}_2 =
\begin{bmatrix}
                    -0.57735&
                    -0.57735&
                    -0.57735&
                    0.0&
                    0.0&
                    0.0
                \\
                    0.57735&
                    0.21132&
                    -0.788675&
                    0.0&
                    0.0&
                    0.0
                \\
                    0.57735&
                    -0.78867&
                    0.21132&
                    0.0&
                    0.0&
                    0.0
                \\
                    0.0&
                    0.0&
                    0.0&
                    1.0&
                    0.0&
                    0.0
            \\
                    0.0&
                    0.0&
                    0.0&
                    0.0&
                    1.0&
                    0.0
                \\
                    0.0&
                    0.0&
                    0.0&
                    0.0&
                    0.0&
                    1.0
\end{bmatrix}
\\ \nonumber
& \mathbf{R}_3 =
\begin{bmatrix}
                    -0.40824&
                    -0.40824&
                    -0.40824&
                    -0.40824&
                    -0.40824&
                    -0.40824&
                \\
                    0.40824&
                    0.71835&
                    -0.28164&
                    -0.28164&
                    -0.28164&
                    -0.28164&
                \\
                    0.40824&
                    -0.28164&
                    0.71835&
                    -0.28164&
                    -0.28164&
                    -0.28164&
                \\
                    0.40824&
                    -0.28164&
                    -0.28164&
                    0.71835&
                    -0.28164&
                    -0.28164&
\\
                    0.40824&
                    -0.28164&
                    -0.28164&
                    -0.28164&
                    0.71835&
                    -0.28164&
\\
                    0.40824&
                    -0.28164&
                    -0.28164&
                    -0.28164&
                    -0.28164&
                    0.71835&
\end{bmatrix}
\end{eqnarray}

To connect the minima, we used points defined with the following parameters

\begin{eqnarray}
\mathbf{p}_1 &=& [1,1,1,1,1,1] \quad \boldsymbol{\sigma}_1=[0.2,0.2,0.2,0.2,0.2,0.2]\\\nonumber
\mathbf{p}_2 &=& [1,1,1,-1,-1,-1] \quad \boldsymbol{\sigma}_2=[0.2,0.2,0.2,0.2,0.2,0.2]\\\nonumber
\mathbf{p}_3 &=& [-1,-1,-1,-1,-1,-1] \quad \boldsymbol{\sigma}_3=[0.2,0.2,0.2,0.2,0.2,0.2]
\end{eqnarray}

To introduce barriers between the minima, gaussian functions were introduced on top of points with the same parameters as the points and height $h=1$.

The total potential is then scaled with a $\alpha=4$.

\section{\label{SI-sec:lj}LJ-38}

As for the toy models, also in the case of LJ-38, ITRE $c_I(t)$
assume a higher value than $c_s(t)$ in the initial part of the trajectory, and it decreases after that a almost steady equilibrium has been achieved. To emphasize once again this behavior, we have illustrated in Fig.~\sref{fig:lj_38_c_t} $c(t)$ calculated with both methods for all the simulations of the LJ-38 system.

\begin{figure}
    \centering
    \includegraphics[width=\textwidth]{figures/c_t_LJ_comparison.pdf}
    \caption{Behavior of $c_I(t)$ and $c_s(t)$ for the LJ-38 system, in full and dashed lines respectively. As for the Langevin particle system, $c_I(t)$ is larger than $c_s(t)$ at early stages of the trajectory, and is smaller once that the simulations reached a quasi-stationary regime. }
    \label{fig:lj_38_c_t}
\end{figure}